\documentclass[pra,preprint,tightenlines,showpacs]{revtex4}
\usepackage{epsfig}

\begin{document}

\title{Muon pair creation from positronium in a circularly polarized laser field}

\author{Carsten M\"uller, Karen Z. Hatsagortsyan, and Christoph H. Keitel}
\affiliation{ Max-Planck-Institut f\"ur Kernphysik, 
Saupfercheckweg 1, D-69117 Heidelberg, Germany}

\date{\today}

\begin{abstract}
We study elementary particle reactions that result from the interaction of an atomic 
system with a very intense laser wave of circular polarization. As a specific example, 
we calculate the rate 
for the laser-driven reaction $e^+e^- \to \mu^+\mu^-$, where the electron and positron
originate from a positronium atom or, alternatively, from a nonrelativistic $e^+e^-$ 
plasma. We distinguish accordingly between the coherent and incoherent channels of the 
process. Apart from numerical calculations, we derive by analytical means compact formulas 
for the corresponding reaction rates. The rate for the coherent channel in a laser 
field of circular polarization is shown to be damped because of the destructive 
interference of the partial waves that constitute the positronium ground-state wave packet. Conditions for the observation of the process via the dominant incoherent channel in a
circularly polarized field are pointed out.
\end{abstract}

%\pacs{13.66.De (Lepton production in e-e+ interactions), 41.75.Jv (Laser-driven acceleration), 
%36.10.Dr (Positronium, muonium, muonic atoms and molecules)}

\maketitle

% body of paper begins here

\section{Introduction}
The interaction of electrons and atoms with laser radiation is intensively and 
successfully being studied for many years now. However, due to a rapid technological 
progress, the high-power laser systems available today can generate peak intensities 
up to 10$^{22}$ W/cm$^2$ in the range of near-optical infrared frequencies \cite{Las}, 
and a further increase can be expected within the next few years \cite{1026}. 
Consequently, the ponderomotive energy of an $e^-$ (or $e^+$) inside such a laser wave 
is of the order of 1 GeV, which is far beyond the typical energetic range of atomic 
physics but rather reaches the energy scale characteristic for elementary particle 
physics. If electrons or an $e^-$ and $e^+$ collide at such high energies, then 
particle reactions like heavy lepton-pair creation or hadron production can occur. 
This indicates that there might be a way to merge laser physics with high-energy physics
\cite{HEP1,HEP2}. Similar efforts are being undertaken with respect to laser physics and 
nuclear physics \cite{nuc,fusion,Mocken,Thomas}. 

The high-energy process of photon-induced $e^+e^-$ pair creation by a projectile particle 
colliding with an intense laser beam has already been investigated before, both 
experimentally \cite{SLAC} and theoretically \cite{we,Reiss}. An essential ingredient to 
these studies is the ultrarelativistic energy of the incoming particle. In its rest frame, 
the doppler-shifted laser frequency and field strength are considerably enhanced. As a 
consequence, the projectile actually faces an x-ray beam of near-critical intensity. 

Instead, in the present paper we study a situation where elementary particle reactions 
arise from the interaction of a strong laser field with a nonrelativistic atomic system. 
To this end, we suppose that a positronium (Ps) atom is brought into an intense laser 
wave. We note that the lifetimes of ortho-Ps ($\sim 10^{-7}$ sec) and para-Ps 
($\sim 10^{-10}$ sec) are much longer than the typical duration of a strong laser pulse.
Due to the equal masses of its constituents, the dynamical response of the positronium 
to the electromagnetic forces exerted by the laser field is rather unique \cite{Ps}: The 
laser's linearly polarized electric field leads to an antiparallel oscillatory motion 
of the particles in the transverse direction, while the magnetic Lorentz force causes an 
identical ponderomotive drift motion along the laser propagation direction. This 
leads to periodic $e^+e^-$ (re)collisions (see, in particular, Fig.\,1 in Ref.\,\cite{Ps}).
If the energy of the relative $e^+e^-$ motion is large enough, then in these {\it coherent}
collisions \cite{coll} particle reactions can occur. Thus, we shall study high-energy 
processes induced by $e^+e^-$ annihilation resulting from a laser-driven Ps atom. 
Considering the case of a circularly polarized laser field we will find as a main result, 
however, that the various partial waves that constitute the Ps ground-state interfere 
destructively, which causes a heavy suppression of the coherent reaction rate. 
This quantum effect can be related to the classical trajectories of the colliding
particles in the laser field. Furthermore, when the characteristic size of these
trajectories (or the size of the spreading particle wave packets) exceeds the 
interatomic distance, then collisions between particles originating from different 
Ps atoms will come into play, which opens the incoherent channel of the process. 
Surprisingly it turns out that, in a circularly polarized laser field, the incoherent 
channel is dominant as the interference in the coherent channel is destructive.  
In order to study the incoherent process we replace the Ps atom by a nonrelativistic 
$e^+e^-$ plasma. In this situation, exclusively incoherent $e^+e^-$ collisions occur. 

It should be stressed that in the described setup the $e^+e^-$ collision energy is basically 
determined by the kinetic energy $\sim mc^2\xi$ contained in the transversal motion of the particles, 
which is considerably smaller than the ponderomotive energy $\sim mc^2\xi^2$ mentioned above. 
Here, $mc^2$ is the electron rest energy and $\xi = ea/mc^2$ denotes the so-called laser intensity 
parameter with the electron charge $-e$ and the laser's vector potential $a$. For the 
highest intensities attainable at present $\xi$ is of order 10$^2$. In this respect,
the underlying laser acceleration of the particles is considerably different from the usual 
laser acceleration techniques, since the latter try to extract the ponderomotive energy gain 
along the laser propagation direction. Nevertheless, the energetic thresholds for muon or pion 
production might be within reach. We further notice, that the principal difficulties 
of laser acceleration implied by the Lawson-Woodward theorem (see, e.g., Ref. \cite{LW}) are 
completely absent here since the $e^+$ and $e^-$ collide {\it inside} the laser wave.

Against this background, we consider the specific process $e^+e^-\to\mu^+\mu^-$ 
which is one of the most fundamental in high-energy physics. Its cross section sets the scale
for all $e^+e^-$ annihilation cross sections \cite{PS}. For example, at high energies one 
has $\sigma_{e^+e^-\to\mbox{\tiny hadrons}} \approx 4\sigma_{e^+e^-\to\mu^+\mu^-}$,
where $\sigma_{e^+e^-\to\mbox{\tiny hadrons}}$ denotes the total cross section for the
production of any number of strongly interacting particles \cite{hadrons}. 
The threshold energy for the reaction $e^+e^-\to\mu^+\mu^-$ amounts to $2Mc^2$ in the 
field-free case, where $M$ denotes the muon mass. According to the above, a naive estimate 
thus suggests that a laser intensity corresponding to $\xi\approx M/m\approx 200$ is 
required to produce a muon pair in a laser-driven $e^+e^-$ collision. This value is reached, 
e.g., for a linearly polarized laser beam of $3.8\times 10^{22}$ W/cm$^2$ intensity and 1 eV 
photon energy. 

To the best of our knowledge, the process $e^+e^-\to\mu^+\mu^-$ in a laser field has not
been considered before. The most closely related article treats the laser-assisted Bhabha 
scattering $e^+e^-\to e^+e^-$ \cite{ee+}. In Ref.\,\cite{ee+} the low-intensity case 
(i.e., $\xi\ll 1$) is analysed in detail with the emphasis lying on the resonances that 
can occur in the scattering cross-section due to the interaction of the leptons with the 
background laser field. We will come back to this point later. Another similar process, 
that has found the interest of several authors, is the M{\o}ller scattering 
$e^-e^-\to e^-e^-$ in a laser field (see \cite{ee1,ee2,ee3,ee4} and references therein). 

The paper is organized as follows. In Sec.\,II we develop a formalism that allows us to 
calculate the rate for the reaction $\mbox{Ps}\to\mu^+\mu^-$ in a strong laser field. 
Our treatment will be based upon the Volkov solutions to the Dirac equation. Afterwards
we analyse in detail the reaction kinematics. Here we show in particular that the 
minimal laser intensity parameter required is indeed given by $\xi_{\rm min}=M/m$ 
[cf. Eq.\,(\ref{ximin})]. Further, the kinematical analysis will help us to derive a compact 
formula that gives an approximation to the total reaction rate and displays its main dependences 
[cf. Eq.\,(\ref{tildeRPs2})]. In Sec.\,III we present our (numerical) results on the total and 
differential production rates and compare them with the known cross section 
for the field-free process $e^+e^-\to\mu^+\mu^-$. Furthermore, we briefly consider the related 
process of muon pair production by a superstrong laser wave interacting with a nonrelativistic 
$e^+e^-$ plasma. In this situation the interference effect does not play a role. We finish with 
a conclusion.

\section{Theoretical framework}

\subsection{Transition amplitude and reaction rate}
We calculate the rate for positronium decay into muons in a strong laser field, i.e., the 
rate for the laser-driven process ${\rm Ps}\to\mu^+\mu^-$. We assume a photon energy of about 
$1\, {\rm eV}$ and a laser intensity parameter of order $M/m\sim 200$ or larger \cite{Thomson}. 
For mathematical simplicity, the laser field is taken to be a monochromatic, plane wave of circular polarization with the classical four-potential \cite{notation}
\begin{eqnarray}
\label{A}
A^\mu(x) = a_1^\mu\cos(kx)+a_2^\mu\sin(kx).
\end{eqnarray}
As usual, $A^\mu$ is assumed to be adiabatically switched on and off in the remote past and the 
distant future, respectively. In Eq.\,(\ref{A}), $k^\mu=\omega(1,0,0,1)$ is the wave four-vector 
and $a_{1,2}^\mu$ are constant four-vectors chosen as $a_1^\mu=(0,a,0,0)$ and $a_2^\mu=(0,0,a,0)$ 
with $a$ denoting the amplitude of the vector-potential. From now on we use relativistic units 
($\hbar=c=1$), except where otherwise stated. We notice that in the circularly polarized 
laser field (\ref{A}) the $e^+$ and $e^-$ are permanently colliding since, according to the classical
equations of motion, they are co-rotating in the polarization plane. 

The Ps atom is assumed to be initially at rest and in its ground state. In a usual field theoretic 
formalism \cite{PS}, this bound initial state can be expressed as a superposition of products of 
free states $\psi_{p_\pm}$ for the electron and positron with definite momenta 
$\mbox{\boldmath$p$}_\pm = \pm\mbox{\boldmath$p$}$. The superposition is weighted by the probability
amplitude $\tilde\Phi(\mbox{\boldmath$p$})$ for finding a particular value of $\mbox{\boldmath$p$}$.
Note that this amplitude is just the Compton profile of the Ps ground state (i.e., the Fourier 
transform of its wave function) and $\mbox{\boldmath$p$}$ can be viewed as the relative momentum of the 
electron-positron two-body system (i.e., as the momentum of an effective particle of reduced mass $m/2$).
When submitted to the strong laser field ($\xi\gtrsim 200$) the Ps atom will instantaneously 
be ionized, and the dynamics of the ionized $e^-$ and $e^+$ will be governed by the laser field, which predominates over the influence of the Coulomb interaction between the particles. Therefore, in the 
spirit of the strong-field approximation theories \cite{SFA}, we may replace the free leptonic states $\psi_{p_\pm}$ by laser-dressed Volkov states \cite{Vol,LL}. Within this framework, the amplitude for 
the laser-driven process ${\rm Ps}\to\mu^+\mu^-$ can be written as
\begin{eqnarray}
\label{SPs}
\mathcal{S}_{{\rm Ps}\to\mu^+\mu^-} = {1\over\sqrt{V}}\int{d^3p\over(2\pi)^3}\,
           \tilde\Phi(\mbox{\boldmath$p$})\, \mathcal{S}_{e^+e^-\to\mu^+\mu^-}
\end{eqnarray}
with a normalization volume $V$ and 
\begin{eqnarray}
\label{See}
\mathcal{S}_{e^+e^-\to\mu^+\mu^-} = -{\rm i}\alpha_{\rm f}\int d^4x \int d^4y
\overline\Psi_{p_+,s_+}(x)\gamma^\mu\Psi_{p_-,s_-}(x) D_{\mu\nu}(x-y)
\overline\Psi_{P_-,S_-}(y)\gamma^\nu\Psi_{P_+,S_+}(y)
\end{eqnarray}
being the amplitude for the process $e^+e^-\to\mu^+\mu^-$ in a laser wave (cf. Fig.\,1).
In Eq.\,(\ref{See}), $\alpha_{\rm f}$ denotes the finestructure constant,
\begin{eqnarray}
D_{\mu\nu}(x-y) = \int {d^4q\over(2\pi)^4}
                  {{\rm e}^{{\rm i}q\cdot(x-y)}\over q^2}g^{\mu\nu}
\end{eqnarray}
is the free photon propagator \cite{prop,prop2}, and the laser-dressed states 
for the electron and positron are given by \cite{Vol,LL}
\begin{eqnarray}
\label{Vol}
\Psi_{p_\pm,s_\pm}(x)= \sqrt{m\over p_\pm^0}
\left(1\pm {e\slash\!\!\!k\slash\!\!\!\!A\over 2(kp_\pm)}\right)
u_{p_\pm,s_\pm}\,{\rm e}^{{\rm i}f^{(\pm)}}
\end{eqnarray}
with
\begin{eqnarray*}
f^{\scriptscriptstyle{(\pm)}} = 
\pm (q_\pm x) + {e(p_\pm a_1)\over (kp_\pm)}\sin(kx) - {e(p_\pm a_2)\over (kp_\pm)}\cos(kx).
\end{eqnarray*}
In Eq.\,(\ref{Vol}), $p_\pm$ are the initial free four-momenta of the electron and positron 
(outside the laser field), $s_\pm$ denote the particle spin states, the $u_{p_\pm,s_\pm}$ are 
free Dirac spinors 
\cite{BD}, and 
\begin{eqnarray}
\label{q}
q_\pm^\mu = p_\pm^\mu +{e^2a^2\over 2(kp_\pm)}k^\mu
\end{eqnarray}
are the effective four-momenta of the particles in the laser field
\cite{LL}. Note that Eq.\,(\ref{q}) implies 
$\mbox{\boldmath$q$}_\pm^\perp = \mbox{\boldmath$p$}_\pm^\perp$ and thus 
$\mbox{\boldmath$q$}_+^\perp + \mbox{\boldmath$q$}_-^\perp = 0$,
where the label $\perp$ denotes the momentum component that is perpendicular to the laser 
propagation direction. The corresponding effective mass reads
$ m_*^2 = q_\pm^2 = (1 + \xi^2)m^2$
with the dimensionless laser intensity parameter
\begin{eqnarray}
\label{xi}
\xi = {ea\over m}.
\end{eqnarray} 
Like free states, the Volkov states in Eq.\,(\ref{Vol}) are normalized to a $\delta$-function 
in $p_\pm$ space \cite{LL,SZ}.
Analogous expressions hold for the Volkov states $\Psi_{P_\pm,S_\pm}$,
the free momenta $P_\pm$, the spin states $S_\pm$,
the effective momenta $Q_\pm^\mu$, the effective mass $M_\ast=M(1+\Xi^2)^{1/2}$,
and the intensity parameter $\Xi=ea/M$ of the muons. Note that the
amplitude (\ref{See}) fully accounts for the interaction of 
the leptons with the laser field, while their interaction with
the QED vacuum is taken into account to lowest order.
Similar approaches have been used for the theoretical description
of laser-assisted $e^+e^-$ \cite{ee+} and $e^-e^-$ \cite{ee1,ee2,ee3,ee4} scattering.

By the standard procedure of using the generating function of the
Bessel functions \cite{AS}, one can perform the space-time
integrations in Eq.\,(\ref{See}) to get
\begin{eqnarray}
\label{See1}
\mathcal{S}_{e^+e^-\to\mu^+\mu^-} &=& -{\rm i} (2\pi)^4 \alpha_{\rm f}
{m\over\sqrt{p_+^0 p_-^0}}{M\over\sqrt{P_+^0 P_-^0}}
\int {d^4q\over q^2} \sum_{n,N} \mathcal{M}^\mu(p_+,p_-|n) \mathcal{M}_\mu(P_+,P_-|N) \nonumber\\
& & \times\ \delta(q_+ + q_- -q -nk)\, \delta(Q_+ + Q_- -q -Nk)
\end{eqnarray}
with the electronic spinor-matrix product
\begin{eqnarray}
\label{M}
\mathcal{M}^\mu(p_+,p_-|n) &=& 
\bar u_{p_+,s_+}\bigg\{\left(\gamma^\mu - 
{e^2a^2 k^\mu\slash\!\!\!k\over 2(kp_+)(kp_-)}\right)b_n^0 \nonumber\\
& &\ \ \ \ \ \ \ \ \,
+\left({e\gamma^\mu\slash\!\!\!k\slash\!\!\!a_1\over 2(kp_+)} -
{e\slash\!\!\!a_1\slash\!\!\!k\gamma^\mu\over 2(kp_-)}\right)b_n^+ \nonumber\\
& &\ \ \ \ \ \ \ \ \, 
+\left({e\gamma^\mu\slash\!\!\!k\slash\!\!\!a_2\over 2(kp_+)} -
{e\slash\!\!\!a_2\slash\!\!\!k\gamma^\mu\over 2(kp_-)}\right)b_n^-
\bigg\} u_{p_-,s_-}
\end{eqnarray}
and a corresponding expression $\mathcal{M}_\mu(P_+,P_-|N)$ for the muons. 
The coefficients in Eq.\,(\ref{M}) are given by 
\begin{eqnarray}
\label{coeff1}
b_n^0 &=& J_n(\alpha)\,{\rm e}^{-{\rm i}n\varphi_0} \nonumber \\
b_n^+ &=& {1\over 2}\,
\left[ J_{n-1}(\alpha){\rm e}^{-{\rm i}(n-1)\varphi_0} + 
       J_{n+1}(\alpha){\rm e}^{-{\rm i}(n+1)\varphi_0} \right] \nonumber\\
b_n^- &=& {1\over 2{\rm i}}
\left[ J_{n-1}(\alpha){\rm e}^{-{\rm i}(n-1)\varphi_0} - 
       J_{n+1}(\alpha){\rm e}^{-{\rm i}(n+1)\varphi_0} \right]
\end{eqnarray}
with
$\alpha = \sqrt{\alpha_1^2+\alpha_2^2}$, $\varphi_0 =
\arccos(\alpha_1/\alpha)=\arcsin(\alpha_2/\alpha)$,
and
\begin{eqnarray}
\label{alpha}
\alpha_j = {e(a_jp_-)\over (kp_-)} - {e(a_jp_+)\over (kp_+)}
\end{eqnarray}
for $j=1,2$. As is expressed by the energy-momentum conserving 
$\delta$-function at the first vertex, the integer number $n$ in
Eq.\,(\ref{See1}) 
counts the laser photons that are emitted (if $n>0$) or absorbed
(if $n<0$) by the electron and positron. Similarly, $N$ is the 
number of laser photons emitted (if $N<0$) or absorbed (if $N>0$)
by the muons. Denoting the total number of absorbed laser photons 
by $r:=N-n$ and integrating over the virtual photon momentum yields
\begin{eqnarray}
\label{See2}
\mathcal{S}_{e^+e^-\to\mu^+\mu^-} &=& -{\rm i} (2\pi)^4 \alpha_{\rm f}
{m\over\sqrt{p_+^0 p_-^0}}{M\over\sqrt{P_+^0 P_-^0}} 
\sum_{n,r} \mathcal{M}^\mu(p_+,p_-|n) \mathcal{M}_\mu(P_+,P_-|n+r)\nonumber\\
& &\times\
{\delta(q_+ + q_- - Q_+ - Q_- + rk)\over (q_+ + q_- - nk)^2}.
\end{eqnarray}
In general, the denominator $(q_+ + q_- - nk)^2$ in Eq.\,(\ref{See2}) could become 
zero. By way of a renormalization procedure, such mathematical singularities can be transformed 
into physical resonances that appear in the production process \cite{ee+,ee1,ee2,ee3}. One can easily 
see, however, that in the present situation, due to the large value of the laser intensity parameter 
and the nonrelativistic electron and positron momenta $p_\pm$, one is always far off resonance \cite{res}. Namely, on the one hand we have
$$ (q_+ + q_- - nk)^2 = 2m_\ast^2+2(q_+q_-)-2n(kp_+)-2n(kp_-)
  \approx 4m_\ast^2 - 4n\omega m $$
which becomes zero for 
\begin{eqnarray}
\label{nres}
n_{\rm res}\approx \xi^2{m\over\omega} \sim 10^{10}.
\end{eqnarray}
On the other hand, the Bessel functions $J_n(\alpha)$, that 
enter the production amplitude through the coefficients in 
Eq.\,(\ref{coeff1}), practically vanish unless $\alpha\gtrsim n$.
Since $q_+^\mu\approx q_-^\mu$ and $m_\ast,q_\perp\ll q_z$,
the argument approximately equals
$$ \alpha \approx 
 2{ea\over\omega}{|\mbox{\boldmath$q$}_\perp|\over q_0-q_z} 
 \approx 4\xi{m\over\omega}
 {|\mbox{\boldmath$q$}_\perp| q_z\over m_\ast^2}, $$
where we have dropped the particle labels $\pm$.
Now, $|\mbox{\boldmath$q$}_\perp| = |\mbox{\boldmath$p$}_\perp|\sim m\alpha_{\rm f}$ 
and $q_z\approx m\xi^2/2$. Hence, 
\begin{eqnarray}
\label{nmax}
 \alpha \sim \alpha_{\rm f}\xi {m\over\omega} \sim 10^6
\end{eqnarray} 
which, according to Eq.\,(\ref{nres}), is orders of magnitude smaller than would be required 
for a resonance to occur.

The above argument can be further exploited. From Eq.\,(\ref{nmax}) we know that the main 
contribution to the production amplitude comes
from photon numbers $n$ with $|n|\lesssim 10^6$. But for those
numbers we have to a very good approximation 
$$ (q_+ + q_- - nk)^2 \approx (q_+ + q_-)^2 $$
which, thus, can be pulled out of the sum in Eq.\,(\ref{See2}):
\begin{eqnarray}
\label{See3}
\mathcal{S}_{e^+e^-\to\mu^+\mu^-} &\approx& -{\rm i} (2\pi)^4 \alpha_{\rm f}
{m\over\sqrt{p_+^0 p_-^0}}{M\over\sqrt{P_+^0 P_-^0}} 
{1\over (q_+ + q_-)^2} 
\sum_{n,r} \mathcal{M}^\mu(p_+,p_-|n) \mathcal{M}_\mu(P_+,P_-|n+r)\nonumber\\
& &\times \ \delta(q_+ + q_- - Q_+ - Q_- + rk).
\end{eqnarray}
The summation over $n$ can now be performed analytically by virtue 
of Graf's addition theorem \cite{AS} with the result
\begin{eqnarray}
\label{uvw}
\sum_n \mathcal{M}^\mu(p_+,p_-|n)\mathcal{M}_\mu(P_+,P_-|n+r) &=&
J_r u^\mu U_\mu + K_r^+ (u^\mu V_\mu + v^\mu U_\mu)
+ K_r^- (u^\mu W_\mu + w^\mu U_\mu) \nonumber\\ 
& & +\ L_r^+ v^\mu V_\mu
+ M_r (v^\mu W_\mu + w^\mu V_\mu) + L_r^- w^\mu W_\mu.
\end{eqnarray}
Here we have used the abbreviations
\begin{eqnarray}
\label{spinor-matrix-products}
u^\mu &=& \bar u_{p_+,s_+}\left(\gamma^\mu - 
{e^2a^2 k^\mu\slash\!\!\!k\over 2(kp_+)(kp_-)}\right)u_{p_-,s_-} \nonumber\\
v^\mu &=& \bar u_{p_+,s_+}
\left({e\gamma^\mu\slash\!\!\!k\slash\!\!\!a_1\over 2(kp_+)} -
{e\slash\!\!\!a_1\slash\!\!\!k\gamma^\mu\over 2(kp_-)}\right)
u_{p_-,s_-} \nonumber \\
w^\mu &=& \bar u_{p_+,s_+}
\left({e\gamma^\mu\slash\!\!\!k\slash\!\!\!a_2\over 2(kp_+)} -
{e\slash\!\!\!a_2\slash\!\!\!k\gamma^\mu\over 2(kp_-)}\right)
u_{p_-,s_-}
\end{eqnarray}
and similarly $U_\mu, V_\mu$, and $W_\mu$ for the muons. 
The coefficients in Eq.\,(\ref{uvw}) read
\begin{eqnarray}
\label{coeff2}
J_r       &=& J_r(\delta) \epsilon^r \nonumber\\
K_r^+ &=& {1\over 2}\left[ J_{r-1}(\delta) \epsilon^{r-1}
              + J_{r+1}(\delta) \epsilon^{r+1} \right]\nonumber \\
K_r^- &=& {{\rm i}\over 2}\left[ J_{r-1}(\delta) \epsilon^{r-1}
              - J_{r+1}(\delta) \epsilon^{r+1} \right]\nonumber \\
L_r^\pm &=& {1\over 4}\left[ 2J_r(\delta)\epsilon^r
        \pm J_{r-2}(\delta) \epsilon^{r-2}
        \pm J_{r+2}(\delta) \epsilon^{r+2} \right]\nonumber \\
M_r &=& {{\rm i}\over 4}\left[ J_{r-2}(\delta) \epsilon^{r-2}
              - J_{r+2}(\delta) \epsilon^{r+2} \right]
\end{eqnarray}
with 
$$\gamma = \beta-\alpha {\rm e}^{{\rm i}(\varphi_0-\eta_0)}\,,\ \
  \delta = |\gamma|\,,\ \ \mbox{and}\ \ 
  \epsilon = {\gamma\over\delta} {\rm e}^{{\rm i}\eta_0} $$
where $\beta$ and $\eta_0$ are the muonic quantities that
correspond to $\alpha$ and $\varphi_0$. We will see later that 
$\gamma\approx\beta$ since $\alpha \ll \beta$ for the
typical parameters. An insignificant overall phase factor of 
${\rm e}^{{\rm i}r\eta_0}$ can be dropped in Eq.\,(\ref{coeff2}).

Now we come back to the reaction ${\rm Ps}\to\mu^+\mu^-$. In order 
to obtain the corresponding amplitude we have, according to Eq.\,(\ref{SPs}), 
to multiply Eq.\,(\ref{See3}) by the Compton profile 
$\tilde\Phi(\mbox{\boldmath$p$})$ of the positronium ground
state and integrate over the relative momentum $\mbox{\boldmath$p$}$.
It turns out that this integration is a very difficult task that can
only be done in an approximate way:

First, within the momentum range given by $\tilde\Phi(\mbox{\boldmath$p$})$
the electronic spinor-matrix products in Eq.\,(\ref{spinor-matrix-products})
are practically constant (on the 1\% level since 
$|\mbox{\boldmath$p$}|/m\sim\alpha_{\rm f}$) and can therefore be pulled out of 
the integration. The same holds for the kinematic factors $p_\pm^0\approx m$,
$(q_++q_-)^2\approx 4m_\ast^2$, and the energy-momentum conserving 
$\delta$-function \cite{delta}. Hence, we are left with integrals of the form
\begin{eqnarray}
\label{int}
 \bar{J_r} = {1\over\sqrt{V}}\int {d^3p\over (2\pi)^3}\, 
             \tilde\Phi(\mbox{\boldmath$p$}) J_r(\delta)\, {\rm e}^{{\rm i}r\chi}
\end{eqnarray}
where 
$\chi=\arctan[\alpha\sin(\varphi_0-\eta_0)/(\alpha\cos(\varphi_0-\eta_0)-\beta)]$
such that ${\rm exp}({\rm i}\chi)=\gamma/\delta$. The highly oscillating factor 
${\rm exp}({\rm i}r\chi)$ leads to a very small value of $\bar{J_r}$. The
oscillatory damping of the amplitude is due to a destructive interference of the
various partial waves within the Ps wave packet. Classically, this interference effect 
can be related to the extended motion of the $e^+$ and $e^-$ in the polarization plane 
of the laser. Therefore, the mean impact parameter of the $e^+e^-$ collisions is much 
larger than the initial Ps size and the resulting $\mu^+\mu^-$ production amplitude 
is suppressed. The laborious evaluation of the integral (\ref{int}) is performed in
the appendix. The result is 
\begin{eqnarray}
\label{barJr}
 \bar{J_r} \approx - {\sqrt{2}\over\pi^{3/2}a_0^{3/2}\sqrt{V}}\, 
 \left({\omega\over m\alpha_{\rm f}\xi}\right)^2\beta^{1/3}\,J_r(\beta).
 \end{eqnarray}
with the Ps radius $a_0=2/\alpha_{\rm f} m$. The damping factor can also be 
written as $(\omega/m\alpha_{\rm f}\xi)^2 = (\pi a_0/\lambda\xi)^2\sim 10^{-12}$. 
Note that $2\lambda\xi$ gives the average impact parameter of the $e^+e^-$ 
collisions since, according to their classical trajectories, the particles 
co-rotate in the polarization plane on opposite sides of a circle of radius 
$\lambda\xi$. However, the classical picture suggests that the process 
probability is proportional to $(a_0/\lambda\xi)^2$. Instead, this factor is 
contained in the process amplitude such that the probability scales as 
$(a_0/\lambda\xi)^4$. This indicates that the damping factor is truely of 
quantum mechanical origin. 

The square of the amplitude reads
\begin{eqnarray}
\label{Ssq}
|\mathcal{S}_{{\rm Ps}\to\mu^+\mu^-}|^2 &=& (2\pi)^4 \alpha_{\rm f}^2
{m^2\over p_+^0 p_-^0}{M^2\over P_+^0 P_-^0} 
{1\over (q_+ + q_-)^4} 
\sum_r \left|\overline{\sum_n} \mathcal{M}^\mu(p_+,p_-|n) \mathcal{M}_\mu(P_+,P_-|n+r) \right|^2\nonumber\\
& &\times\ \delta(q_+ + q_- - Q_+ - Q_- + rk) VT\,,
\end{eqnarray}
where $\overline{\sum}_n$ indicates the sum over $n$ in Eq.\,(\ref{uvw}) averaged 
over the Ps ground state, as described above, $p_\pm$ ($q_\pm$) are to be understood 
as some typical values of the electron and positron (effective) momenta, and the 
factors of volume $V$ and time $T$ come, as usual, from the square of the $\delta$-function. 
Note that the energy-momentum conserving $\delta$-function, in 
particular, implies $\mbox{\boldmath$Q$}_+^\perp + \mbox{\boldmath$Q$}_-^\perp =0$.
From Eq.\,(\ref{Ssq}) we get the total reaction rate by averaging over the initial 
spin states, summing over the final spin states, and integrating over the final momenta:
\begin{eqnarray}
\label{RPs}
R_{{\rm Ps}\to\mu^+\mu^-} = {1\over T}
\int {d^3 P_+\over (2\pi)^3} \int {d^3 P_-\over (2\pi)^3}\, 
{1\over 4} \sum_{s_\pm,S_\pm}|\mathcal{S}_{{\rm Ps}\to\mu^+\mu^-}|^2.
\end{eqnarray}
In the next but one subsection we derive a compact analytical formula that gives an 
estimate for the muon production rate (\ref{RPs}). But before, we analyse in some 
detail the reaction kinematics.

\subsection{Kinematical considerations}
In the following we provide estimates for the minimal ($r_{\rm min}$) and 
the typical ($\bar r$) photon numbers that are net-absorbed during the
production process. From the latter we also find the typical momenta of the 
created muons.

A lower bound on $r$ can be derived from the equation
\begin{eqnarray}
 (q_+ + q_- + rk)^2 = 2(q_++q_-+rk)\cdot Q_\pm
\end{eqnarray}
that follows from the energy-momentum conservation condition expressed by the 
$\delta$ function in Eq.\,(\ref{Ssq}). Setting $Q_r\equiv q^+_0 + q^-_0 + r\omega$ 
and $q_r\equiv q_z^+ +q_z^- + r\omega$, this can be rewritten as
\begin{eqnarray}
\label{cos}
\cos\theta_{Q_\pm} = {2Q_rQ_\pm^0-(Q_r^2-q_r^2)\over  2q_r|\mbox{\boldmath$Q$}_\pm|}
\end{eqnarray}
with the polar angle
$\theta_{Q_\pm}=\angle(\mbox{\boldmath$k$},\mbox{\boldmath$Q$}_\pm)$.
Demanding $\cos^2\theta_{Q_\pm}\le 1$, we get
\begin{eqnarray}
\label{Erange}
 \left| Q_\pm^0 - {Q_r\over 2} \right| \le {q_r\over 2}
 \left( 1 - {4M_\ast^2\over Q_r^2-q_r^2} \right)^{1/2}.
\end{eqnarray}
Hence, it is required that 
$ 4M_\ast^2 \le Q_r^2-q_r^2 = (q_+ + q_- +rk)^2 $, i.e., the 
laser-dressed collision energy has to exceed twice the laser-dressed muon mass.
Using $(q_+ + q_- +rk)^2 \approx 4m_\ast^2 + 4r\omega m$, we find
\begin{eqnarray}
\label{rmin}
r \gtrsim r_{\rm min} \equiv
{M^2-m^2\over\omega m} \approx {M^2\over\omega m}.
\end{eqnarray} 
This means that, e.g., for $\omega = 1$ eV at least $2\times 10^{10}$ photons have to 
be absorbed from the laser wave for muon production to take place from the initially 
low-energy $e^+e^-$ pair. This number is independent of the laser intensity. 

Assuming a symmetric situation with $Q_+^0\approx Q_-^0$, Eq.\,(\ref{rmin}) 
implies that the minimal muon energy is approximately given by 
\begin{eqnarray}
\label{Emin}
Q_{\rm min}^0 \approx {1\over 2}Q_{r_{\rm min}} 
              \approx {m\over 2}\left( \xi^2 + {M^2\over m^2}\right)
              \approx {M^2\over m}.
\end{eqnarray}
Hence, the muons are typically produced with highly relativistic momenta
such that their dispersion relation approximately reads
$Q_\pm^0\approx |\mbox{\boldmath$Q$}_\pm|$. Furthermore, they are emitted 
roughly along the laser propagation direction (note that $Q_z\gg Q_\perp$ since
$q_r\approx m\xi^2+r\omega$, $Q_r\approx 2m+m\xi^2+r\omega$ so that 
$q_r\approx Q_r$). More precisely, by solving Eq.\,(\ref{cos}) for $Q_\pm^ 0$ 
we find that the polar emission angle satisfies the relation
\begin{eqnarray}
\label{cos1}
\cos\theta_{Q_\pm} \ge {Q_r\over q_r}
 \left[ 1-{(Q_r^2-q_r^2)^2\over 4Q_r^2M_\ast^2} \right]^{1/2}
 \approx \left({1-\frac{4m^2}{M^2_*}}\right)^{1/2}.
\end{eqnarray}
For these reasons one can say that the muon kinematics is similar to that of the laser photons. This "photon-like" nature of the muons results from the fact that they are essentially produced by a huge number of laser photons whose total energy, according to Eq.\,(\ref{Emin}), exceeds the initial nonrelativistic energy of the $e^-$ and 
$e^+$ by orders of magnitude.

The typical number of absorbed laser photons can be estimated by exploiting 
the properties of the Bessel function $J_r(\beta)$ in Eq.\,(\ref{barJr}). 
To this end, let us again assume a symmetric situation, which allows us to drop 
the particle indices $\pm$ in what follows. The energy-conservation condition 
then can approximately be written as $2Q_0\approx 2q_0+r\omega$. Because of
the photon-like muon momenta this can be expressed as 
\begin{eqnarray}
2Q_z\left( 1+{Q_\perp^2+M_\ast^2\over 2Q_z^2}\right) \approx
2q_z\left( 1+{q_\perp^2+m_\ast^2\over 2q_z^2}\right) + r\omega,
\end{eqnarray}
where $q_z\approx q_0\approx m\xi^2/2$. Applying the momentum conservation 
condition $2Q_z \approx 2q_z + r\omega$, we thus get
\begin{eqnarray}
\label{star}
{Q_\perp^2+M_\ast^2\over Q_z} \approx {q_\perp^2+m_\ast^2\over q_z}.
\end{eqnarray}
Now, let $r_0\equiv q_0/\omega$ and $\ell\equiv r/r_0$. Then, again by the momentum-conservation condition, $Q_z\approx (1+\ell/2)q_z$. Hence, 
Eq.\,(\ref{star}) implies 
\begin{eqnarray}
\label{Qperp}
 Q_\perp \approx \left[ \left(1+{\ell\over 2}\right)m_\ast^2
 - M_\ast^2 \right]^{1/2}
\end{eqnarray}
where $m_\ast \gg q_\perp$ was used. Thus, the argument of the Bessel functions in
Eq.\,(\ref{barJr}) approximately equals
\begin{eqnarray}
\label{beta}
  \beta 
  \approx 2{ea\over\omega}{Q_\perp\over Q_0-Q_z}
  \approx 4\xi{m\over\omega}
  {Q_\perp Q_z\over Q_\perp^2+M_\ast^2}
  \approx 2\xi{Q_\perp\over\omega}
  \approx 2\xi{m\over\omega}\left({r\omega\over m}-{M^2\over m^2}\right)^{1/2}.
\end{eqnarray}
We note that, according to Eq.\,(\ref{rmin}), the expression under the square root
on the right-hand side of Eq.\,(\ref{beta}) is positive.
By the properties of the Bessel functions \cite{AS}, the typical number of absorbed
laser photons is expected to be determined by the condition $\beta\approx r$. 
This yields
\begin{eqnarray}
\label{barr}
r \sim \bar r \equiv 2\xi^2 {m\over\omega}\left( 1 + \sqrt{1-\kappa^2}\right)
\end{eqnarray}
with $\kappa \equiv M/m\xi$ \cite{rpm}. The corresponding typical muon momenta read
\begin{eqnarray}
\label{barQ}
 {\bar Q}_\perp &\approx& m \xi \left( 1+\sqrt{1-\kappa^2}\right), \nonumber\\
 {\bar Q}_z &\approx& {m\over 2}\xi^2 
 \left[ 1 + 2\left( 1 + \sqrt{1-\kappa^2}\right)\right].
\end{eqnarray}
Using the Eq.\,(\ref{q}) between the effective and the free four-momenta and the
relations (\ref{kp}) below, we find for the typical values of the muon momenta 
after the interaction with the laser field
\begin{eqnarray}
\label{barP}
{\bar P}_\perp &=& {\bar Q}_\perp 
               \,\approx\, m\xi\left( 1+\sqrt{1-\kappa^2} \right), \nonumber\\
{\bar P}_z &\approx& {\bar Q}_z - {m\over 2}\xi^2 
            \,\approx\, m\xi^2\left( 1 + \sqrt{1-\kappa^2} \right).
\end{eqnarray}
For example, for $\xi = 250$ and $\omega = 1$ eV we have $\bar r\approx 10^{11}$, 
${\bar Q}_\perp = {\bar P}_\perp \approx 2 M$, ${\bar Q_z}\approx 620 M$, and 
${\bar P_z}\approx 470 M$. From Eqs.\,(\ref{barr}) and (\ref{barP}) we see that
the typical final energy of the muon pair satisfies the relation 
$2\bar P_0\approx\bar r\omega$, which reflects the law of energy conservation
after the laser has been switched off. Equation\,(\ref{barr}) also implies that 
the minimal intensity parameter required for the process to have a significant 
probability (i.e., to be able to fulfill $r\approx \beta(r)$) amounts to 
\begin{eqnarray}
\label{ximin}
 \xi_{\rm min} = {M\over m}
\end{eqnarray}
which agrees with our earlier naive estimate.

We notice that the partial rate for muon production by the absorption
of $r=r_{\rm min}$ photons is zero. Namely, according to Eqs.\,(\ref{Qperp}) and
(\ref{beta}), for $\ell = r_{\rm min}/r_0\approx 2M^2/m^2\xi^2$ the transverse
muon momentum and with it the argument of the Bessel functions practically vanish.
According to the above, the partial rate reaches a maximum at $r\sim \bar{r}$, 
which at $\xi=\xi_{\rm min}$ is twice as large as the minimal number of photons: 
$\bar{r}=2r_{\rm min}$.

It is interesting to observe that the typical muon momenta in Eq.\,(\ref{barP}) can 
be interpreted by employing a classical simple man's model of the creation process. 
In the classical picture, the threshold value of the laser intensity (\ref{ximin})  
corresponds to the situation when, in the center-of-mass frame of the $e^+e^-$ 
system, the kinetic energy is large enough to create muons at rest. I.e., denoting 
the laser phase by $\tau \equiv \omega(t-z)$, we have
\begin{equation}
\label{P0}
P^{\prime}_\perp(\tau_0) = P^{\prime}_z(\tau_0) = 0
\end{equation}
at the creation phase $\tau_0$, where the prime indicates the center-of-mass frame. 
The classical equation of motion for a muon in a laser field with the initial 
condition (\ref{P0}) has the solution
\begin{eqnarray}
P^{\prime}_\perp(\tau) &=& eA(\tau)-eA(\tau_0)\,, \nonumber \\
P^{\prime}_z(\tau) &=& \frac{e^2}{2M}\left[A(\tau)-A(\tau_0)\right]^2.
\label{P_t}
\end{eqnarray}
Consequently, after the interaction with the laser field the muon momenta equal
\begin{eqnarray}
P^\prime_\perp &=& m\xi\,, \nonumber \\
P^\prime_z &=& \frac{m^2}{2M}\xi^2.
\label{P_inf}
\end{eqnarray}
Due to the $e^+e^-$ longitudinal drift motion in the laser field, the
relative velocity between the center-of-mass frame and the lab frame amounts to
$v_{\rm rel} = q_z/q_0 = \xi^2/(2+\xi^2)$ [see Eq.\,(\ref{q})]. The Lorentz 
transformation to the lab frame thus yields
\begin{eqnarray}
P_\perp &=& m\xi\,, \nonumber \\
P_z &=& {M\over 2}\frac{\xi^2}{\sqrt{1+\xi^2}}+\frac{m^2}{2M}\xi^2\sqrt{1+\xi^2}.
\label{P_lab}
\end{eqnarray}
The latter coincides with the typical muon momenta at the threshold 
$\xi=\xi_{\rm min}$ given by the quantum theory: $\bar P_z\approx M^2/m$ and 
$\bar P_\perp\approx M$ [see Eq.\,(\ref{barP})]. The typical number of absorbed 
photons is determined by the muon final energy: $\bar{r}\omega= 2(P_0-m)\approx 2M^2/m$, 
which is in agreement with Eq.\,(\ref{barr}).

Our simple man's model can also explain a peculiarity in the angular distribution of 
the muons (see Fig. 4 in Sec. III.\,A). Since $P_z\gg P_{\bot}$, the muons move in a 
narrow cone with the axis parallel to the laser propagation direction, but at very 
small angles the angular spectrum has a dip. The dark region in the angular distribution 
occurs because the muons, although having been created with zero transverse momentum in 
the laser field, acquire a nonvanishing transverse momentum after switching off the 
laser field [see Eqs.\,(\ref{P0}) and (\ref{P_t})].

\subsection{An approximative formula for the total rate}
Equation (\ref{RPs}) for the total rate of the reaction ${\rm Ps}\to\mu^-\mu^+$, 
although looking rather innocent, is actually quite involved and can be evaluated only 
numerically. Therefore it is desirable to find, by analytical means, an approximation 
to Eq.\,(\ref{RPs}) that displays its main physical content. To this end, we consider 
the contribution to the total rate stemming from the first term on the right-hand side 
of Eq.\,(\ref{uvw}). From our numerical calculations we learn that this
term gives by far the main contribution ($\sim 90\,\%$). Thus, we need to calculate
\begin{eqnarray}
\label{tildeRPs}
\tilde R_{{\rm Ps}\to\mu^+\mu^-} &=& {\alpha_{\rm f}^2 a_0\over 2^7\pi^5}
\left({\omega\over\xi}\right)^4 
{m^2\over p_+^0p_-^0}{M^2\over (q_+ + q_-)^4}
\int {d^3 P_+\over P_+^0} \int {d^3 P_-\over P_-^0} \nonumber\\
& & \times\, \sum_r [J_r(\beta)]^2 \beta^{2/3}
\sum_{s_\pm,S_\pm}|u^\mu U_\mu|^2 \, \delta(q_+ + q_- - Q_+ - Q_- + rk).
\end{eqnarray}
With the help of the $\delta$-function and the relations \cite{LL}
$$ {d^3P_\pm\over P_\pm^0} = {d^3Q_\pm\over Q_\pm^0}\,,\ \  d^3Q_- = 
 |\mbox{\boldmath$Q$}_-| Q_-^0 dQ_-^0d\cos\theta_{Q_-} d\phi_{Q_-} $$
we can integrate over $d^3Q_+$ and $dQ_-^0$ to find
\begin{eqnarray}
\label{tildeRPs1}
\tilde R_{{\rm Ps}\to\mu^+\mu^-} \approx {\alpha_{\rm f}^2 a_0\over 2^{11}\pi^5}
\left({\omega\over\xi}\right)^4 {M^2\over m_\ast^4}
\int d\phi_{Q_-} \int d\cos\theta_{Q_-}
\sum_r [J_r(\beta)]^2 \beta^{2/3} \sum_{s_\pm,S_\pm}|u^\mu U_\mu|^2
\end{eqnarray}
where the relations $p_\pm^0 \approx m$,  
$(q_++q_-)^2\approx 4 m_\ast^2$, and 
$|\mbox{\boldmath$Q$}_-|\approx Q_+^0$ have been used. The spin sum in 
Eq.\,(\ref{tildeRPs1}) can be converted in the usual way into a product of 
two traces:
\begin{eqnarray}
\label{T}
\mathcal{T}_{uU,uU} &:=& 
\sum_{s_\pm,S_\pm} \left| u^\mu U_\mu \right|^2 \nonumber\\
 &=& {\rm Tr}\left\{ \left( \gamma^\mu - 
{e^2a^2 k^\mu\slash\!\!\!k\over 2(kp_+)(kp_-)} \right)
{\slash\!\!\!p_-+m\over 2m} \left( \gamma^\nu - 
{e^2a^2 k^\nu\slash\!\!\!k\over 2(kp_+)(kp_-)} \right)
{\slash\!\!\!p_+-m\over 2m} \right\} \nonumber\\
&\times& {\rm Tr}\left\{ \left(\gamma_\mu - 
{e^2a^2 k_\mu\slash\!\!\!k\over 2(kP_+)(kP_-)} \right)
{\slash\!\!\!\!P_-+M\over 2M} \left( \gamma_\nu - 
{e^2a^2 k_\nu\slash\!\!\!k\over 2(kP_+)(kP_-)} \right)
{\slash\!\!\!\!P_+-M\over 2M} \right\}.
\end{eqnarray}
We want to find some typical value of $\mathcal{T}_{uU,uU}$. The standard 
trace technology yields
\begin{eqnarray}
\label{Trace1}
\mathcal{T}_{uU,uU} &\approx& 
{2\over m^2M^2}\left[ (p_-P_-)(p_+P_+) + (p_-P_+)(p_+P_-) \right]
+ {2(P_+P_-)\over M^2} + {2(p_+p_-)\over m^2} \nonumber\\
& & - {2\xi\Xi\over mM} 
\left[ (p_-P_-)+(p_-P_+)+(p_+P_+)+(p_+P_-)-2(P_+P_-)-2(p_+p_-)\right]
\nonumber\\
& & + 4\xi^2 + 4\Xi^2 + 4\xi^2\Xi^2 + 4.
\end{eqnarray}
Here we have used the (remarkable) relations
\begin{eqnarray}
\label{kp}
\omega m \approx (kp_+) \approx (kp_-) \approx (kP_+) \approx (kP_-)
\end{eqnarray}
that hold to a good approximation since
\begin{eqnarray}
(kP) = (kQ) \approx
\omega{Q_\perp^2+M_\ast^2\over 2Q_z} \approx
\omega{q_\perp^2+m_\ast^2\over 2q_z} \approx (kq) = (kp)
\end{eqnarray}
by Eq.\,(\ref{star}). With $(p_+p_-)\approx m^2$, 
$(p_\pm P_\pm)\approx mP_0$, and $(P_+P_-)\approx M^2 + 2P_\perp^2$
the expression in Eq.\,(\ref{Trace1}) becomes
\begin{eqnarray}
\label{T2}
\mathcal{T}_{uU,uU} \approx
{4\over M^2} (P_0 - m\xi^2)^2 + 8\xi^2{P_\perp^2\over M^2}
\end{eqnarray}
with $P_0$ and $P_\perp$ denoting some characteristic values of the muonic 
energies and transversal momenta that, by Eq.\,(\ref{barP}), amount to
$P_0 \approx 2m\xi^2$ and $P_\perp \approx 2m\xi$ at $\xi\gg\xi_{\rm min}$. 
This leads to the desired typical value of
\begin{eqnarray}
\label{T3}
\mathcal{T}_{uU,uU} \approx 36 \xi^4 {m^2\over M^2},
\end{eqnarray}
which can be pulled out of the integration in Eq.\,(\ref{tildeRPs1}). 
We proceed by performing the further approximations
\begin{eqnarray}
\label{approx}
 \int d\phi_{Q_-} \approx 2\pi\,,\ \ 
   \int d\cos\theta_{Q_-} \approx {\theta_{\rm max}^2\over 2}\,,\ \
   \beta^{2/3}\approx {\bar r}^{\,2/3}\,,\ \
   \sum_r [J_r(\beta(r))]^2 \approx 1\,,
\end{eqnarray}
where, according to Eq.\,(\ref{cos1}), the maximum polar emission angle 
is given by 
\begin{eqnarray}
\label{thetamax}
\theta_{\rm max} \approx {2m\over M_\ast}.
\end{eqnarray}
Putting all pieces together, we arrive at the handy formula
\begin{eqnarray}
\label{tildeRPs2}
\tilde R_{{\rm Ps}\to\mu^+\mu^-} \approx 
{3^2\over 2^6\pi^4} {\alpha_{\rm f}^2\over \xi^2} 
\left({\omega^2\over m_\ast M_\ast}\right)^2 
%\left[ {2m\xi^2\over\omega}\left( 1+\sqrt{1-{M^2\over m^2\xi^2}} \right) \right]^{2/3}
\left( {4m\xi^2\over\omega} \right)^{2/3}
{1\over r_e}
\end{eqnarray}
where $r_e$ denotes the classical electron radius \cite{hbarc}. 
Equation (\ref{tildeRPs2}) is the desired analytical estimate for the rate
of laser-driven Ps decay into muons.
%From our numerical calculations we learn that, for $\xi\gtrsim\xi_{\rm min}$,
%Eq.\,(\ref{tildeRPs2}) gives indeed a very reasonable estimate of the total 
%reaction rate (\ref{RPs}), if we include an additional factor of 2 accounting 
%for the (many) terms that are not included in the approximative treatment above. 
%Thus, for the total rate of muon production by 
%laser-driven positronium annihilation we have found the handy formula
%\begin{eqnarray}
%\label{RPs3}
%R_{{\rm Ps}\to\mu^-\mu^+} \approx 
%{5\over 2^{10}\pi^3} {\alpha_{\rm f}\over \xi^6}\left({\omega\over mc^2}\right)^4
%\left( {m_\ast^2\over M_\ast^2}-{1\over 5} \right) {c\over\dbar},
%\end{eqnarray}
%where $\dbar$ denotes the Compton wave length of the electron. 

We notice that Eq.\,(\ref{tildeRPs2}) can also be represented in the form
\begin{eqnarray}
\tilde R_{{\rm Ps}\to\mu^+\mu^-} \approx \frac{\sigma}{a_0^3} \left(\frac{a_0}{\lambda\xi}\right)^4 \left( {4m\xi^2\over\omega} \right)^{2/3}.
\label{rate_Ps}
\end{eqnarray}
%with the factor $(a_0/\lambda\xi)^4$ arising from the destructive interference 
%within the Ps wave packet. 
Here $\sigma = (9/8)(\alpha_{\rm f}^2/M_\ast^2)$ stands for 
the process cross section, which for $\xi\gg\xi_{\rm min}$ becomes
\begin{eqnarray}
\label{sigma}
\sigma \approx {9\over 8}{r_e^2\over\xi^2}.
\end{eqnarray}
Recalling the simple man's model and using the electron energy in the center-of-mass frame $p^{\prime}_0\approx m\xi$, we can infer that Eq.\,(\ref{rate_Ps}) is based on 
the process cross section 
\begin{eqnarray}
\sigma\sim \frac{r_e^2}{\gamma^2},
\label{cross-section}
\end{eqnarray}
with $\gamma=p^{\prime}_0/m\approx\xi$ being the electron gamma-factor in the center-of-mass 
frame. Equation (\ref{cross-section}) is in accordance with the 
known field-free cross section for muon production in $e^+e^-$ collisions 
[see Eq.\,(\ref{field-free})]. 

If the interaction volume contains $N$ positronium atoms, then the rate will increase 
correspondingly:
 \begin{eqnarray}
R_{{\rm Ps}\to\mu^+\mu^-} ^{(N)}= N R_{{\rm Ps}\to\mu^+\mu^-}.
\label{rate_Ps_N}
\end{eqnarray}
Here we have assumed that each Ps atom independently creates a muon pair, i.e. there 
is no interference between electrons (positrons) stemming from different Ps atoms. 
The latter is the case when the electron wave-packets from different atoms do not 
overlap, i.e. when $\lambda\xi \ll n^{-1/3}$, where $n$ is the Ps density. Otherwise, 
when the spatial extension of the electron wave-packet is large, then the gas of Ps 
atoms transforms into an $e^+e^-$ plasma. Against this background, in the following we 
will denote the rate in Eq.\,(\ref{rate_Ps_N}) resp.\,(\ref{RPs}) as 
the {\it coherent} rate for muon production since the colliding $e^+$ and $e^-$ 
originate from one and the same Ps atom. In contrast to that, the rate for muon
creation from an $e^+e^-$ plasma, that might have been formed from an initial 
Ps gas, will be refered to as {\it incoherent} rate. In this situation, electrons and
positrons from different Ps atoms can collide which gives a total number of $N^2$
incoherent collisions. In the next section we will present our results both on the 
coherent and the incoherent channel of muon production.

\section{Results and discussion}

\subsection{Muon pair creation by a laser-driven Ps atom: The coherent process}

Based on Eq.\,(\ref{RPs}), we have numerically calculated the coherent rate for 
$\mu^+\mu^-$ pair creation from a single Ps atom submitted to a strong laser field 
of circular polarization. 
The laser frequency has been taken to be $\omega = 1$ eV, throughout. 
For the laser intensity parameter we have chosen the three different 
values $\xi = 250$, 500, and 1000. The corresponding laser intensities 
amount to $1.1\times 10^{23}$ W/cm$^2$, $4.5\times 10^{23}$ W/cm$^2$, and 
$1.8\times 10^{24}$ W/cm$^2$, respectively. 

In Fig.\,2, the dependence of the total muon creation rate on the laser 
intensity parameter is shown. For the intensity parameters under 
consideration we find production rates of $1.0\times 10^{-15}$ sec$^{-1}$ 
($\xi = 250$), $1.6\times 10^{-16}$ sec$^{-1}$ ($\xi = 500$), and 
$1.1\times 10^{-17}$ sec$^{-1}$ ($\xi = 1000$). 
The analytical approximation (\ref{tildeRPs2}) overestimates these numbers, 
but is still in rather good agreement with them (cf. Fig.\,2). The reason
for the overestimation, in particular for $\xi\approx\xi_{\rm min}$, are the 
rather large values of $P_0$ and $P_\perp$ used in Eq.\,(\ref{T2}). 
A total production rate of 10$^{-15}$ sec$^{-1}$ means that in a finite laser 
pulse of femtosecond duration the probability to create a muon pair from a single Ps
atom is of order 10$^{-30}$. We notice that a typical laser focal volume 
will contain only one Ps atom on average since the highest positronium densities 
achievable at present are on the order of 10$^8$ cm$^{-3}$ \cite{density}.
However, proposals to reach a Ps density of $10^{14}$ cm$^{-3}$ \cite{perez} or even 
a Ps Bose-Einstein condensate of $10^{18}$ cm$^{-3}$ \cite{cassidy} are being considered.
The above numbers of created muons seem too small to be 
experimentally accessible. Clearly, the main reason for the smallness of the coherent 
reaction rate lies in the damping factor $(a_0/\lambda\xi)^4\sim 10^{-26}$ 
[cf. Eq.\,(\ref{barJr})]. The latter results from the destructive interference of the
partial waves constituting the Ps wave packet in the laser field, which makes the 
recollision of an $e^+$ and $e^-$ from the same Ps atom highly unlikely. From a 
simplified, classical point of view, the reason for the damping lies in the large 
collisional impact parameter of order $\lambda\xi$ that is due to the equal handed 
rotations of the particles in the laser polarization plane \cite{spread}.

In Fig.\,3, the partial production rates with respect to the number $r$ of 
absorbed laser photons are shown (i.e., the contributions to the total rate 
stemming from a net absorption of $r$ laser photons in the production process). 
The photon number is given in units of $r_0 \approx m\xi^2/2\omega$, 
which amounts to $1.6\times 10^{10}$ ($\xi = 250$), $6.4\times 10^{10}$ 
($\xi = 500$), and $2.6\times 10^{11}$ ($\xi = 1000$), respectively. 
The shape of the curves in Fig.\,3 can be understood with 
the help of the kinematical analysis in Sec.\,II.\,B. 
First, according to Eq.\,(\ref{rmin}), the minimal number of laser photons 
required from kinematical constraints amounts to $r_{\rm min} = 2.2\times 10^{10}$,
independent of the laser intensity. If we express this number 
with respect to the respective values of $r_0$, then we get $r_{\rm min}/r_0 = 1.4$ 
($\xi = 250$), $0.3$ ($\xi = 500$), and $0.1$ ($\xi = 1000$). 
The partial reaction rate for $r = r_{\rm min}$ is always zero, as can be seen in 
Fig.\,3. Further, in agreement with Eq.\,(\ref{barr}), the curves exhibit maxima 
at $r\approx \bar r$; for $\xi=250$ the maximum is located at $\bar r/r_0\approx 6$, 
while for $\xi = 500$ and 1000 we have $\bar r/r_0\approx 8$. This feature reflects 
the mathematical properties of the Bessel functions and is correctly predicted by 
our simple man's model. 

Figure 4 shows the angular distribution $dR/d\theta_Q$ for one of the produced
muons. For symmetry reasons, the spectra for the muon $\mu^-$ and the antimuon
$\mu^+$ are identical. The differential rate is expressed with respect to the
polar angle $\theta_Q=\angle(\mbox{\boldmath$k$},\mbox{\boldmath$Q$})$ of the
effective muon momentum. The value of the laser intensity parameter is chosen
to be $\xi = 250$. One can see that the muon is emitted into a very narrow 
angular range starting from $1.3\times 10^{-3}\,{\rm rad}\approx 0.07^\circ$
and extending to $4.7\times 10^{-3}\,{\rm rad}\approx 0.27^\circ$, which is 
in agreement with Eq.\,(\ref{thetamax}). In other words, as has already been
mentioned before, the muon moves practically parallel to the propagation 
direction of the laser beam. The occurence of a minimal emission angle 
arises from the fact that, according to Eq.\,(\ref{beta}), the argument of the 
Bessel function is proportional to the transverse momentum component $Q_\perp$, 
which itself is proportional to $\sin\theta_Q\approx \theta_Q$. Thus, the emission
angle cannot be too small because otherwise the Bessel function will vanish.
Since $P_\perp = Q_\perp$, a dark angular region also exists in the
angular spectrum $dR/d\theta_P$ with respect to the muon momentum outside the
laser beam. An alternative explanation of this phenomenon in terms of a
simple man's model has been given at the end of Sec.\,II.\,B.

\subsection{Comparison with the field-free process $e^+e^-\to\mu^+\mu^-$}
In this subsection we want to draw a comparison between the coherent muon creation 
from a laser-driven Ps atom and the corresponding field-free process 
$e^+e^-\to\mu^+\mu^-$. In the high-energy limit, the cross section for this
reaction reads \cite{PS}
\begin{eqnarray}
\label{field-free}
 \sigma_{\rm free} = 
 {4\pi\over 3} {m^2\over s} r_e^2
\end{eqnarray}
where $\sqrt{s}\gg 2M$ denotes the collision energy. In the case with laser field, 
the square root of the quantity
\begin{eqnarray}
\label{cmenergy}
 (q_++q_-+\bar rk)^2 \approx 4m_\ast^2 + 4\bar r\omega m
 \approx 20m^2\xi^2
\end{eqnarray}
can be regarded as some average collision energy. For the $\xi$-values considered 
in this paper, the laser-dressed collision energy is thus about 1 GeV. Hence, the 
reference cross section in Eq.\,(\ref{field-free}) to compare with should be taken 
at $\sqrt{s}\approx 1$ GeV, where its value is about 100 nbarn.
To transform this cross section into a reaction rate, we have to multipy by the 
incident particle flux. In collider experiments, instead of the incoming flux, 
the luminosity is more commonly used. When a beam of $N_+$ positrons collides 
at high energy with a beam of $N_-$ electrons, then the luminosity is given by
\begin{eqnarray}
\label{L}
 L = {N_+N_-\over UA}
\end{eqnarray} 
where $U$ is the circumference of the collider ring and $A$ is the beam cross 
sectional area at the collision point. To make the comparison with a single 
laser-driven Ps atom, we use $N_\pm = 1$ along with the typical values 
$U\approx 10^3$ m and $A\approx 10^{-5}$ cm$^2$. Note that the corresponding 
mean impact parameter $\sim\sqrt{A}$ of the field-free collision is of the same
order of magnitude as the electron-positron spatial separation $\sim\lambda\xi$ 
in the laser wave. The resulting luminosity $L\sim 10^{11}$ cm$^{-2}$sec$^{-1}$ 
leads to a muon creation rate of 10$^{-20}$ sec$^{-1}$. This number is considerably 
smaller than the production rates we found in Sec.\,III.\,A. However, in a real 
collider experiment one has bunches of $N_\pm\sim 10^{10}$ particles leading to
much higher luminosities and reaction rates, of course.

\subsection{Muon production from laser-plasma interaction: The incoherent process}
We have seen in Sec.\,III.\,A that the large electron-positron wave-packet size 
during the motion in the laser field suppresses the coherent reaction rate
dramatically. To reduce this size and achieve $e^+e^-$ collisions at microscopic
impact parameters, one can think of employing different, more complicated field 
configurations \cite{coll}. Otherwise, the muon creation from Ps atoms will be 
dominated by the incoherent production channel introduced at the end of Sec.\,II.\,C. 
The latter coincides with the process of muon
creation, when a low-energy $e^+e^-$ plasma interacts with a strong laser 
beam. We have redone our calculation for this situation, i.e., for a free,
initially nonrelativistic $e^+$ and $e^-$ that collide in a strong laser 
field and create a $\mu^+\mu^-$ pair by annihilation. Our results on the partial 
production rates for reasonable experimental parameters (see below) are shown in 
Fig.\,5. The shape of the curves is similar to those in Fig.\,3. 

From Eqs.\,(\ref{tildeRPs2}) and (\ref{barJr}) one can infer that the total rate 
for muon creation from laser-plasma interaction approximately reads
\begin{eqnarray}
R \approx\frac{9}{16\pi}\frac{m^2r_e^2}{M_\ast^2}\frac{N_+N_-}{V_{\rm int}}
\label{rate_plasma}
\end{eqnarray}
where $N_\pm$ denotes the number of electrons and positrons in the interaction 
volume $V_{\rm int}$, which is given by the laser focal volume. From 
Eq.\,(\ref{rate_plasma}) we can estimate the total number of produced muons 
$N_{\mu}=R\tau N_s$ during the interaction with $N_s$ laser shots, each single 
shot having a pulse duration of $\tau$. When plugging in some reasonable numbers: 
$\tau=100$ fs, $V_{\rm int}\approx (10\lambda)^3\approx 10^{-9}$ cm$^3$, 
and assuming that the presently achievable number of positrons $N_+\approx 10^7\approx N_-$ \cite{positron_source} can be compressed into the interaction volume
or, alternatively, is created via a newly emerging laser-based technique \cite{wilks},
then we get $N_{\mu}\approx1$ muon production event during $N_s=10^{10}$ shots. 
This number, being based on rather optimistic experimental parameters 
(especially concerning the positron compression), indicates 
that the realisation of the incoherent muon creation process might be not inhibitory
difficult, but still it will be very hard with modern experimental techniques.

\section{Conclusion}
In this paper we have studied $\mu^+\mu^-$ production by $e^+e^-$ annihilation 
from a laser-driven Ps atom. To this end, a calculational framework has been developed
where the initial bound state is described as a superposition of Volkov states
weighted by the Compton profile of the Ps ground state. By virtue of the interaction
with the QED vacuum, which is treated in the first order of perturbation theory,
this initial state can decay into a laser-dressed $\mu^+\mu^-$ pair. Also, the
related process of muon creation by the interaction of a strong laser field with
a low-energy $e^+e^-$ plasma has been examined.

We have found that
the minimal laser intensity required for the process to occur corresponds to 
an intensity parameter of $\xi\approx M/m\approx 200$. In the case of a near-optical 
laser wave of circular polarization, e.g.,  this value is reached for an intensity 
of $7\times 10^{22}$ W/cm$^2$. This means that, starting from a nonrelativistic Ps 
atom or $e^+e^-$ plasma, fundamental particle reactions can be ignited by a 
superstrong laser field of an intensity that is just one order of magnitude larger 
than the highest values available today.

However, in the Ps case, the total production rate resulting from the coherent 
recollisions is extremely small and amounts to about 10$^{-15}$ per second only. 
The strong suppression is caused by a destructive interference of the different 
partial waves constituting the bound initial state in the superintense laser field.
This phenomenon is also expressed by a compact formula for the total rate
that we derived by analytical means. 
As a consequence, in the considered setup the production rate will be dominated
by the muon creation via incoherent $e^+e^-$ scattering, for which the system
of Ps atoms has no advantage compared to an $e^+e^-$ plasma. For the incoherent collisions,
the interference plays no role and the resulting muon creation rate is significantly 
larger than the corresponding rate from the coherent production channel.
Nevertheless, very demanding experimental conditions are required in order to
achieve observable muon yields. 

Finally, we note that similar reaction rates can be expected for laser-driven 
$\pi^+\pi^-$ production from $e^+e^-$ or $\mu^+\mu^-$. A more promising alternative 
within circularly polarized field configurations might be the process 
$e^+e^-\to e^+e^- + e^+e^-$, the field-free cross section of which being several 
orders of magnitude larger than the one for $\mu^+\mu^-$ creation in
Eq.\,(\ref{field-free}) \cite{2ee}.
%The case of the laser-plasma interaction
%is more favorable since the spreading is absent here. Nevertheless, the expected 
%muon yields are still very small. For this reason we conclude that the considered 
%process seems hardly to be observable in an experiment.

\newpage
\renewcommand{\theequation}{A\arabic{equation}}
\setcounter{equation}{0}
\appendix
\begin{center}
  {\bf APPENDIX}
\end{center}
In this appendix we calculate the integral in Eq.\,(\ref{int}). Using cylindrical 
coordinates it reads
\begin{eqnarray}
\label{J1}
 \bar{J_r} = {1\over (2\pi)^3\sqrt{V}}\int_{-\infty}^{+\infty} dp_z 
                  \int_0^{\infty} p_\perp d p_\perp \int_{-\pi}^{+\pi} d\varphi\,
                  \tilde\Phi(\mbox{\boldmath$p$}) J_r(\delta)\, {\rm e}^{{\rm i}r\chi}
\end{eqnarray}
with 
\begin{eqnarray*}
 \delta &=& \left[ \alpha^2 + \beta^2 - 
                   2\alpha\beta\cos(\varphi_0-\eta_0) \right]^{1/2},\\
 \alpha &=& {2m\xi\over\omega}{p_0p_\perp\over p_\perp^2+m^2}\,, \\
 \chi   &=& \arctan\left\{ 
            \alpha\sin(\varphi_0-\eta_0) \over \alpha\cos(\varphi_0-\eta_0) - \beta
            \right\}, \\
 \tilde\Phi(\mbox{\boldmath$p$}) &=&
 {8\sqrt{\pi} a_0^{3/2} \over [1+(a_0p_\perp)^2+(a_0p_z)^2]^2}\,,
\end{eqnarray*}
and $\beta\approx r$ being of order $10^{11}$ for the laser parameters under 
consideration [see Eq.\,(\ref{barr})]. Note that $\varphi_0$ coincides with the 
azimuthal angle $\varphi$. Going over to the variables $a_0p_\perp\to x$, 
$a_0p_z\to z$, and $(\varphi_0-\eta_0)\to\varphi$, Eq.\,(\ref{J1}) reads
\begin{eqnarray}
\label{J2}
 \bar{J_r} = {2\over \pi^{5/2} a_0^{3/2}\sqrt{V}}\int_0^\infty dz 
             \int_0^{\infty} x\,dx \int_{-\pi}^{+\pi} d\varphi\,
             {J_r(\delta)\, {\rm e}^{{\rm i}r\chi}\over (1+x^2+z^2)^2}.
\end{eqnarray}
For brevity, the overall factor $2/(\pi^5 a_0^3 V)^{1/2}$ will be dropped 
in what follows and only restored in the final result [see Eq.\,(\ref{J4})]. 
Since $\alpha\ll\beta$ we can expand the functions $\delta$ and
$\chi$:
\begin{eqnarray*}
\delta &=& \beta - \alpha\cos\varphi + \mathcal{O}\left({\alpha^2\over\beta}\right)\,,
\\
\chi   &=& -{\alpha\over\beta}\sin\varphi + \mathcal{O}\left({\alpha^2\over\beta^2}\cos^2\varphi\right).
\end{eqnarray*}
Note here that, since $\alpha^2/\beta\lesssim 1$ [see Eq.\,(\ref{nmax})], we may 
drop terms of this order in the expansion of $\delta$. Further, 
since the Bessel function $J_r(\delta)$ is exponentially or oscillatorily damped
unless $|r-\delta|\lesssim r^{1/3}$, the main contribution to the integral comes
from the region with $|\alpha\cos\varphi|\lesssim r^{1/3} \sim \beta^{1/3}$. 
Accordingly, terms of order $r(\alpha\cos\varphi/\beta)^2 \lesssim \beta^{-1/3}$ 
may be neglected in the phase $r\chi$. Hence, we obtain
\begin{eqnarray*}
\label{J3}
 \bar{J_r} \approx \int_0^\infty dz \int_0^{\infty} {x\,dx \over (1+x^2+z^2)^2}
             \int_{-\pi}^{+\pi} d\varphi\,
             J_r(\beta-\alpha\cos\varphi)\, {\rm e}^{-{\rm i}\alpha\sin\varphi}.
\end{eqnarray*}
Next we observe that $\bar{J_r} = 2\,{\rm Re}\,I_r$ with
\begin{eqnarray}
\label{I1}
 I_r &\equiv& \int_0^\infty dz \int_0^{\infty} {x\,dx \over (1+x^2+z^2)^2}
         \int_0^{\pi} d\varphi\,
         J_r(\beta-\alpha\cos\varphi)\, {\rm e}^{-{\rm i}\alpha\sin\varphi}
 \nonumber\\
 &=& \int_0^\infty dz \int_0^{\infty} {x\,dx \over (1+x^2+z^2)^2}
  \int_0^{\pi/2} d\varphi\,
  \left[ J_r(\beta-\alpha\cos\varphi) + J_r(\beta+\alpha\cos\varphi) \right]
  \,{\rm e}^{-{\rm i}\alpha\sin\varphi} \nonumber\\
 &\approx& 2 \int_0^\infty dz \int_0^{\infty} {x\,dx \over (1+x^2+z^2)^2}
           \int_0^{\pi/2} d\varphi\,J_r(\beta-\alpha\cos\varphi)
           \,{\rm e}^{-{\rm i}\alpha\sin\varphi}.
\end{eqnarray}
In the last step we exploited the fact that the contributing ranges within
the exponential and oscillatory regions of the Bessel function have a similar 
size. Let now
$$ \zeta = {x\sqrt{x^2+z^2+\epsilon^{-2}}\over x^2+\epsilon^{-2}}\,,
\ \ \rho = \zeta\cos\varphi $$
with $\epsilon\equiv \alpha_{\rm f}/2$, such that $\alpha = \alpha_0\zeta$ 
with $\alpha_0 \equiv 2m\xi/\omega$. Performing the substitution of variables 
$(x,z,\varphi)\to(x,\zeta,\rho)$ we obtain
\begin{eqnarray*}
\label{I2}
I_r &=& 2\int_0^\infty {x^4\,dx\over (x^2+\epsilon^{-2})^3}
\int_{\zeta_{\rm min}}^\infty {\zeta\,d\zeta\over
\sqrt{\zeta^2-{x^2\over x^2+\epsilon^{-2}}}
\left[ \zeta^2-{x^2(\epsilon^{-2}-1)\over(x^2+\epsilon^{-2})^2}\right]^2}
\nonumber\\
& & \times \int_0^\zeta {d\rho\over\sqrt{\zeta^2-\rho^2}}\,J_r(\beta-\alpha_0\rho)\,
{\rm e}^{-{\rm i}\alpha_0\sqrt{\zeta^2-\rho^2}}
\end{eqnarray*}
with $\zeta_{\rm min}=x/\sqrt{x^2+\epsilon^{-2}}$. As mentioned above, the
value of $J_r(\beta-\alpha_0\rho)$ will be exponentially small unless
$\rho\lesssim\rho_0\equiv\beta^{1/3}/\alpha_0$. Therefore, we can approximately 
write
\begin{eqnarray}
\label{I3}
I_r &\approx& 2J_r(\beta)\int_0^\infty {x^4\,dx\over (x^2+\epsilon^{-2})^3}
\int_{\zeta_{\rm min}}^\infty {\zeta\,d\zeta\over
\sqrt{\zeta^2-{x^2\over x^2+\epsilon^{-2}}}
\left[ \zeta^2-{x^2(\epsilon^{-2}-1)\over(x^2+\epsilon^{-2})^2}\right]^2}
\int_0^{\rho_{\rm max}}{d\rho\over\sqrt{\zeta^2-\rho^2}}\,
 {\rm e}^{-{\rm i}\alpha_0\sqrt{\zeta^2-\rho^2}} \nonumber\\
 \,
\end{eqnarray}
with $\rho_{\rm max}\equiv{\rm min}\{\zeta,\rho_0\}$. 

In the following, the integral in Eq.\,(\ref{I3}) will be evaluated in several 
steps. Let us first consider the integral over $\rho$. We show:
\begin{eqnarray}
\label{Izeta}
 I(\zeta) \equiv \int_0^{\rho_{\rm max}} d\rho\,
 {{\rm e}^{-{\rm i}\alpha_0\sqrt{\zeta^2-\rho^2}}\over\sqrt{\zeta^2-\rho^2}}\,
 \approx \left\{\begin{array}{l}
   \displaystyle{ {\pi\over 2}
   \ \ \ \ \ \ \ \ \ \ \ \ \ \ \ \ \ \ \ \ \ \ \ (\zeta \ll \alpha_0^{-1}),}\\
   \\
   \displaystyle{ \sqrt{\pi\over 2}
                  {{\rm e}^{{\rm i}{\pi\over 4}}\over\sqrt{\zeta\alpha_0}}\,
                  {\rm e}^{-{i}\alpha_0\zeta}
                  \ \ \ \ \ (\alpha_0^{-1} \ll \zeta\ll\alpha_0\rho_0^2),}\\
   \\
   \displaystyle{ {\rho_0\over\zeta}\, {\rm e}^{-{i}\alpha_0\zeta}
                  \ \ \ \ \ \ \ \ \ \ \ \ \ \ (\zeta \gg \alpha_0\rho_0^2).}   
                 \end{array}\right.
\end{eqnarray}
Note that, for the parameters of interest, we have $\alpha_0^{-1}\sim 10^{-8}$
and $\rho_0\sim 10^{-5}$ such that $\alpha_0\rho_0^2\sim 10^{-2}$. In the
first range ($\zeta\ll \alpha_0^{-1}$) we get
$$ I(\zeta)\approx \int_0^{\zeta}{d\rho\over\sqrt{\zeta^2-\rho^2}}={\pi\over 2}.$$
The second range we split into two cases. For $\alpha_0^{-1}\ll\zeta < \rho_0$ the
integrand is highly oscillating and the main contribution comes from the boundary
terms: 
\begin{eqnarray*}
I(\zeta) &=& \int_0^\zeta dy\,{{\rm e}^{-{\rm i}\alpha_0y}\over\sqrt{\zeta^2-y^2}}
\approx {1\over {\rm i}\alpha_0\zeta}+ 
\int^\zeta dy\,{{\rm e}^{-{\rm i}\alpha_0y}\over\sqrt{\zeta^2-y^2}}\\
&\approx& {{\rm e}^{-{\rm i}\alpha_0\zeta}\over\sqrt{2\zeta}}
\int_0 dt\,{{\rm e}^{{\rm i}\alpha_0 t}\over\sqrt{t}}
\approx \sqrt{\pi\over 2}
{{\rm e}^{{\rm i}{\pi\over 4}}\over\sqrt{\zeta\alpha_0}}\,
{\rm e}^{-{i}\alpha_0\zeta}\,,
\end{eqnarray*}
where in the first and second steps the substitutions $y=\sqrt{\zeta^2-\rho^2}$
and $t=\zeta-y$ were made. Note here that the contribution from the lower bound
can be neglected since $\alpha_0\zeta\gg 1$.
Similarly, for $\rho_0 < \zeta\ll \alpha_0\rho_0^2$ we can write
\begin{eqnarray*}
I(\zeta) = \int_{\sqrt{\zeta^2-\rho_0^2}}^\zeta dy\,
{{\rm e}^{-{\rm i}\alpha_0y}\over\sqrt{\zeta^2-y^2}}\approx 
{{\rm e}^{-{\rm i}\alpha_0\sqrt{\zeta^2-\rho_0^2}}\over {\rm i}\alpha_0\rho_0}
+ \int^\zeta dy\,{{\rm e}^{-{\rm i}\alpha_0y}\over\sqrt{\zeta^2-y^2}}
\approx \sqrt{\pi\over 2}
{{\rm e}^{{\rm i}{\pi\over 4}}\over\sqrt{\zeta\alpha_0}}\,
{\rm e}^{-{i}\alpha_0\zeta}.
\end{eqnarray*}
Finally, in the third range ($\zeta\gg\alpha_0\rho_0^2$) we obtain
$$ I(\zeta) \approx \int_0^{\rho_0} {d\rho\over\zeta}\,{\rm e}^{-{\rm i}\alpha_0\zeta}
   = {\rho_0\over\zeta}\,{\rm e}^{-{\rm i}\alpha_0\zeta}. $$
This shows Eq.\,(\ref{Izeta}).

Now we continue with the integrations over $dx\,d\zeta$. To this end, let us divide 
the integration range into five regions:
\begin{eqnarray*}
\label{regions}
{\rm I}&:&  \ \ 0\le x\le x_1\,,\ \ \, \zeta_{\rm min}\le\zeta\le\zeta_1 \\
{\rm II}&:& \ \ 0\le x\le x_1\,,\ \ \, \zeta_1\le\zeta\le\zeta_2 \\
{\rm III}&:&\ \ x_1\le x\le x_2\,,\ \zeta_{\rm min}\le\zeta\le\zeta_2 \\
{\rm IV}&:& \ \ 0\le x\le x_2\,,\ \ \, \zeta\ge\zeta_2 \\
{\rm V}&:&  \ \ x\ge x_2\,,\ \ \ \ \ \ \ \ \zeta\ge\zeta_{\rm min}
\end{eqnarray*}
with
$$     x_1 = {1\over\alpha_0\epsilon} \sim 10^{-6}\,,\ \ 
       x_2 = {\alpha_0\rho_0^2\over\epsilon}\sim 1\,,\ \ 
   \zeta_1 = {1\over\alpha_0}\sim 10^{-8}\,,\ \ 
   \zeta_2 = \alpha_0\rho_0^2 \sim 10^{-2}. $$
Note that $\zeta_1$ and $\zeta_2$ coincide with the values of $\zeta_{\rm min}$ 
taken at positions $x_1$ and $x_2$, respectively. Hence, it is easily seen, 
that this division covers the whole range of integration, i.e.:
$$ I_r = I_r^{(I)} + I_r^{(II)} + I_r^{(III)} + I_r^{(IV)} + I_r^{(V)}. $$
Taking into account Eq.\,(\ref{Izeta}), we find in the first range
\begin{eqnarray*}
I_r^{(I)} &\approx& J_r(\beta)\pi \int_0^{(\alpha_0\epsilon)^{-1}}
{x^4\,dx\over\epsilon^{-6}} \int_{x\epsilon}^{\alpha_0^{-1}}
{\zeta\,d\zeta\over\sqrt{\zeta^2-x^2\epsilon^2}
\left[ \zeta^2 - x^2\epsilon^4(\epsilon^{-2}-1) \right]^2}\nonumber\\
&=& J_r(\beta)\pi {\epsilon\over\alpha_0^2}\int_0^1 x^4\,dx
\int_x^1 {\zeta\,d\zeta\over\sqrt{\zeta^2-x^2}
\left( \zeta^2 - x^2 + x^2\epsilon^2 \right)^2}
\end{eqnarray*}
where in the second step the transformations $\alpha_0\epsilon x\to x$,
$\alpha_0\zeta\to\zeta$ have been made. Introducing the variable 
$t=\zeta^2-x^2$, this becomes
\begin{eqnarray}
\label{IrI}
I_r^{(I)} &=& J_r(\beta){\pi\over 2} {\epsilon\over\alpha_0^2}
\int_0^1 x^4\,dx \int_0^{1-x^2}{dt\over\sqrt{t}\,(t+x^2\epsilon^2)^2}
\nonumber\\
&=& J_r(\beta){\pi\over 2}{1\over\alpha_0^2\epsilon^2}
\left\{ \epsilon\int_0^1 {x^2\sqrt{1-x^2}\,dx\over(1-x^2+x^2\epsilon^2)}
+ \int_0^1 x\,{\rm arctan}\left({\sqrt{1-x^2}\over x\epsilon}\right)dx \right\}
\end{eqnarray}
where in the second step formula 1.2.15.13 from Ref.\,\cite{Prudnikov} was used.
Since both integrals in Eq.\,(\ref{IrI}) are of order unity, we conclude that
\begin{eqnarray}
I_r^{(I)} \approx J_r(\beta){\pi\over 2}{1\over\alpha_0^2\epsilon^2}.
\end{eqnarray}
Turning to the second range, we get similar as before
\begin{eqnarray*}
I_r^{(II)} &=& J_r(\beta) \sqrt{2\pi\over\alpha_0}\,{\rm e}^{{\rm i}{\pi\over 4}}
\int_0^{(\alpha_0\epsilon)^{-1}}{x^4\,dx\over\epsilon^{-6}}
\int_{\alpha_0^{-1}}^{\alpha_0\rho_0^2}
{\sqrt{\zeta}\,{\rm e}^{-{\rm i}\alpha_0\zeta}\, d\zeta
\over\sqrt{\zeta^2-x^2\epsilon^2}
\left[ \zeta^2 - x^2\epsilon^4(\epsilon^{-2}-1) \right]^2}\nonumber\\
&=& J_r(\beta)\sqrt{2\pi}\,{\epsilon\over\alpha_0^2}\,
{\rm e}^{{\rm i}{\pi\over 4}}\int_0^1 x^4\,dx \int_1^{\alpha_0^2\rho_0^2}
{\sqrt{\zeta}\, {\rm e}^{-{\rm i}\zeta}\,d\zeta\over\sqrt{\zeta^2-x^2}
\left( \zeta^2 - x^2 + x^2\epsilon^2 \right)^2}.
\end{eqnarray*}
The main contribution to the highly oscillating integral over $\zeta$ comes
from the lower boundary. In this way we find
\begin{eqnarray}
\label{IrII}
I_r^{(II)} &\approx& J_r(\beta)\sqrt{2\pi}\,{\epsilon\over\alpha_0^2}\,
{\rm e}^{{\rm i}{\pi\over 4}}\int_0^1 {x^4\,dx\,
{\rm e}^{-{\rm i}}\over{\rm i}\sqrt{1-x^2}(1-x^2+x^2\epsilon^2)^2}.
\end{eqnarray}
The remaining integral can be done analytically. Its value is
${\pi\over 4}\epsilon^{-3}$ such that we get
\begin{eqnarray}
I_r^{(II)} &\approx& J_r(\beta) \left({\pi\over 2}\right)^{3/2} {{\rm e}^{-{\rm i}{\pi\over 4}}\,{\rm e}^{-{\rm i}}\over\alpha_0^2\epsilon^2}.
\end{eqnarray}
In the third region the integral reads
\begin{eqnarray*}
I_r^{(III)} &=& J_r(\beta) \sqrt{2\pi\over\alpha_0}\,{\rm e}^{{\rm i}{\pi\over 4}}
\int_{(\alpha_0\epsilon)^{-1}}^{\alpha_0\rho_0^2\epsilon^{-1}}
{x^4\,dx\over\epsilon^{-6}}
\int_{x\epsilon}^{\alpha_0\rho_0^2}
{\sqrt{\zeta}\,{\rm e}^{-{\rm i}\alpha_0\zeta}\, d\zeta
\over\sqrt{\zeta^2-x^2\epsilon^2}
\left[ \zeta^2 - x^2\epsilon^4(\epsilon^{-2}-1) \right]^2}\nonumber\\
&=& J_r(\beta)\sqrt{2\pi}\,{\epsilon\over\alpha_0^2}\,
{\rm e}^{{\rm i}{\pi\over 4}}\int_1^{\alpha_0^2\rho_0^2} x^4\,{\rm e}^{-{\rm i}x}\,dx 
\int_0^{\alpha_0^2\rho_0^2-x}
{\sqrt{t+x}\, {\rm e}^{-{\rm i}t}\,dt\over\sqrt{t(t+2x)}
\left[ t(t+2x) + x^2\epsilon^2 \right]^2}
\end{eqnarray*}
where first the same transformations as before were made 
($\alpha_0\epsilon x\to x$, $\alpha_0\zeta\to\zeta$) and then the variable 
$t=\zeta-x$ was introduced. Because of the highly oscillating integrand and the
singularity, the main contribution to the $t$ integration comes from the region
around $t=0$. Thus, we arrive at
\begin{eqnarray}
I_r^{(III)} &\approx& J_r(\beta)\sqrt{2\pi}\,{\epsilon\over\alpha_0^2}\,
{\rm e}^{{\rm i}{\pi\over 4}}\int_1^{\alpha_0^2\rho_0^2} 
{x^4\,{\rm e}^{-{\rm i}x}\,\sqrt{x}\over \sqrt{2x}\,(x\epsilon)^4}\,dx 
\int_0^\infty {{\rm e}^{-{\rm i}t}\over\sqrt{t}}\,dt \nonumber\\
&=& J_r(\beta) {{\rm i}\pi\over\alpha_0^2\epsilon^3}\,
\left({\rm e}^{-{\rm i}\alpha_0^2\rho_0^2}-{\rm e}^{-{\rm i}}\right).
\end{eqnarray}
According to Eq.\,(\ref{Izeta}), in the forth range we have
\begin{eqnarray*}
I_r^{(IV)} &=& 2 J_r(\beta) \rho_0
\int_0^{\alpha_0\rho_0^2\epsilon^{-1}} {x^4\,dx\over\epsilon^{-6}}
\int_{\alpha_0\rho_0^2}^\infty
{{\rm e}^{-{\rm i}\alpha_0\zeta}\, d\zeta\over\sqrt{\zeta^2-x^2\epsilon^2}
\left[ \zeta^2 - x^2\epsilon^4(\epsilon^{-2}-1) \right]^2}\nonumber\\
&=& 2 J_r(\beta) \epsilon\alpha_0\rho_0^3 \int_0^1 x^4\,dx \int_1^\infty
{{\rm e}^{-{\rm i}\alpha_0^2\rho_0^2\zeta}\,d\zeta\over\sqrt{\zeta^2-x^2}
\left( \zeta^2 - x^2 + x^2\epsilon^2 \right)^2}
\end{eqnarray*}
where we have substituted $(\alpha_0\rho_0^2\epsilon^{-1})^{-1}x\to x$ and $(\alpha_0\rho_0^2)^{-1}\zeta\to\zeta$. The highly oscillating integral over 
$\zeta$ can be evaluated with the same methods as before to give
\begin{eqnarray}
I_r^{(IV)} &\approx& 2J_r(\beta)\,{\epsilon\rho_0\over\alpha_0}\,
\int_0^1 {x^4\,dx\,{\rm e}^{-{\rm i}\alpha_0^2\rho_0^2}\over
{\rm i}\sqrt{1-x^2}\,(1-x^2+x^2\epsilon^2)^2}.
\end{eqnarray}
The integral over $x$ we have already encountered in Eq.\,(\ref{IrII}). Hence:
\begin{eqnarray}
I_r^{(IV)} &\approx& J_r(\beta) {\pi\over 2{\rm i}}
{\rho_0\over\alpha_0\epsilon^2}\,{\rm e}^{-{\rm i}\alpha_0^2\rho_0^2}.
\end{eqnarray}
In the fifth range, after the substitution $\epsilon x\to x$, the integral 
can be written as
\begin{eqnarray*}
I_r^{(V)} &\approx& 2 J_r(\beta) \epsilon \rho_0
\int_{\alpha_0\rho_0^2}^\infty {x^4\,dx\over (1+x^2)^3}
\int_{x\over\sqrt{1+x^2}}^\infty {{\rm e}^{-{\rm i}\alpha_0\zeta}\,d\zeta\over
\sqrt{\zeta^2-{x^2\over 1+x^2}}
\left[ \zeta^2-{x^2(1-\epsilon^2)\over(1+x^2)^2}\right]^2}.
\end{eqnarray*}
With the definitions $a=x/\sqrt{1+x^2}$ and $b=x^2(\epsilon^2+x^2)/(1+x^2)^2$,
the $\zeta$ integral can be cast into the form
$$ \int_a^\infty {{\rm e}^{-{\rm i}\alpha_0\zeta}\,d\zeta\over
\sqrt{\zeta^2-a^2}\,(\zeta^2-a^2+b)^2}
\approx {{\rm e}^{-{\rm i}\alpha_0 a}\over\sqrt{2a}\,b^2}
\int_0^\infty {{\rm e}^{-{\rm i}\alpha_0 t}\,dt\over\sqrt{t}}
= \sqrt{\pi\over2\alpha_0}\,{{\rm e}^{-{\rm i}{\pi\over 4}}\over\sqrt{a}\,b^2}
\,{\rm e}^{-{\rm i}\alpha_0 a}$$
where the substitution $t=\zeta-a$ was made and the oscillatory nature
of the integrand exploited. This yields
\begin{eqnarray*}
I_r^{(V)} &\approx& J_r(\beta) \sqrt{2\pi\over\alpha_0}\, \epsilon \rho_0\,
{\rm e}^{-{\rm i}{\pi\over 4}} \int_{\alpha_0\rho_0^2}^\infty
{(1+x^2)^{5/4}\over\sqrt{x}\,(x^2+\epsilon^2)^2}\,
{\rm e}^{-{\rm i}\alpha_0 {x\over\sqrt{1+x^2}}}\,dx.
\end{eqnarray*}
The main contribution to this integral stems from the lower boundary since
for $x\gg 1$, where the oscillations fade away, the integrand is damped by
the power $x^{-13/4}$. For this reason we obtain
\begin{eqnarray}
I_r^{(V)} &\approx& J_r(\beta) \sqrt{2\pi} \,
{{\rm e}^{-{\rm i}{3\pi\over 4}} {\rm e}^{-{\rm i}\alpha_0^2\rho_0^2}\over
\alpha_0^2\epsilon^3}.
\end{eqnarray}
In summary we have shown that
$$ I_r^{(I)} \sim I_r^{(II)} \sim 10^{-12}J_r(\beta)\,,\ \ 
   I_r^{(III)} \sim I_r^{(V)} \sim 10^{-10}J_r(\beta)\,,\ \
   I_r^{(IV)} \sim 10^{-9}J_r(\beta)\,, $$
i.e., the main contribution to the integral (\ref{I3}) comes from region IV.
Consequently, 
\begin{eqnarray}
\label{J4}
\bar{J_r} \approx 2\,{\rm Re}\,I_r^{(IV)} =
-{2\over\pi^{3/2}a_0^{3/2}\sqrt{V}}\, {\beta^{1/3}\over\alpha_0^2\epsilon^2}
\,J_r(\beta)\,\sin\left(\beta^{2/3}\right)
\end{eqnarray}
where we have restored the factor $2/(\pi^5 a_0^3 V)^{1/2}$ that was 
dropped from Eq.\,(\ref{J2}).

We note that Eq.\,(\ref{J4}) introduces a fast oscillating factor 
$\sin^2(\beta^{2/3})$ to the differential reaction rate. Since 
$\beta\sim 2m\xi^2/\omega$ these oscillations depend on the laser intensity 
and frequency [see also Eq.\,(\ref{beta})]. In reality, however, neither of 
these parameters has a definite value in a short laser pulse but spreads 
over a certain range. In an experiment the oscillations are therefore 
averaged out, and we shall do the same: 
\begin{eqnarray}
\label{J5}
\bar{J_r} \approx - {\sqrt{2}\over\pi^{3/2}a_0^{3/2}\sqrt{V}}\, {\beta^{1/3}\over\alpha_0^2\epsilon^2}\,J_r(\beta).
\end{eqnarray}

\newpage
\begin{center}
{\bf FIGURE CAPTIONS}
\end{center}

Fig.\,1. Feynman diagram for the process $e^+e^-\to \mu^+\mu^-$
in a laser field. The arrows are labelled by the free and 
effective momenta of the corresponding particle [cf. Eq.\,(\ref{q})].\\

Fig.\,2. Total rates for the laser-driven reaction 
$\mbox{Ps}\to \mu^+\mu^-$ as a function of the intensity parameter
$\xi$ of the applied laser field. The black squares show the results 
of our numerical calculations based on Eq.\,(\ref{RPs}); the
solid line shows the analytical estimate via Eq.\,(\ref{tildeRPs2}).\\
 
Fig.\,3. Partial rates for the laser-driven reaction 
$\mbox{Ps}\to \mu^+\mu^-$ as a function of the number of absorbed
laser photons $r$ for an intensity parameter of $\xi = 250$ (solid line),
500 (dashed line), and 1000 (dotted line). The latter two curves are
enhanced by factors of 10$^2$ and $5\times 10^3$, respectively. \\
 
Fig.\,4. Angular spectrum with respect to the polar angle of the 
effective momentum for one of the produced muons at $\xi=250$. \\
 
Fig.\,5. Partial rates for the reaction $e^+e^-\to \mu^+\mu^-$ 
from a laser-driven nonrelativistic plasma. It is assumed that 
$N_\pm = 10^7$ particles are contained in the interaction volume  
$V_{\rm int} = 10^{-9}$ cm$^3$. The laser intensity 
parameter is $\xi = 250$ (solid line), 500 (dashed line), and 1000 
(dotted line). The latter two curves are enhanced by factors of 
$2\times 10^2$ and $2\times 10^4$, respectively. \\

\vspace{3cm}
\begin{figure}[h]
\begin{center}
\resizebox{6cm}{!}{\includegraphics{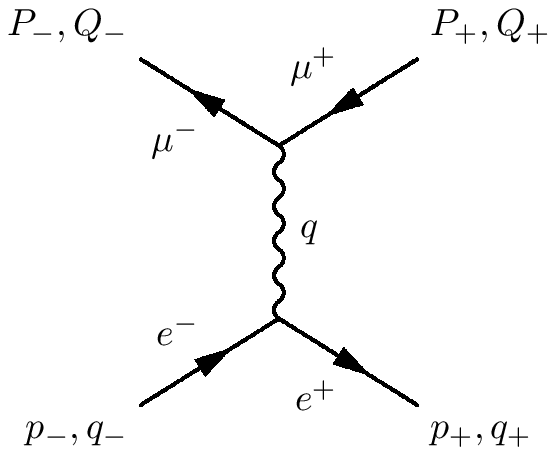}}
\end{center}
\end{figure}

\newpage
\begin{figure}[t]
\begin{center}
\resizebox{13cm}{!}{\includegraphics{muellerfig2.eps}}
\end{center}
\end{figure}

\begin{figure}[b]
\vspace{1cm}
\begin{center}
\resizebox{13cm}{!}{\includegraphics{muellerfig3.eps}}
\end{center}
\end{figure}
 
\newpage
\begin{figure}[t]
\begin{center}
\resizebox{13cm}{!}{\includegraphics{muellerfig4.eps}}
\end{center}
\end{figure}

\begin{figure}[b]
\begin{center}
\resizebox{13cm}{!}{\includegraphics[angle=-90]{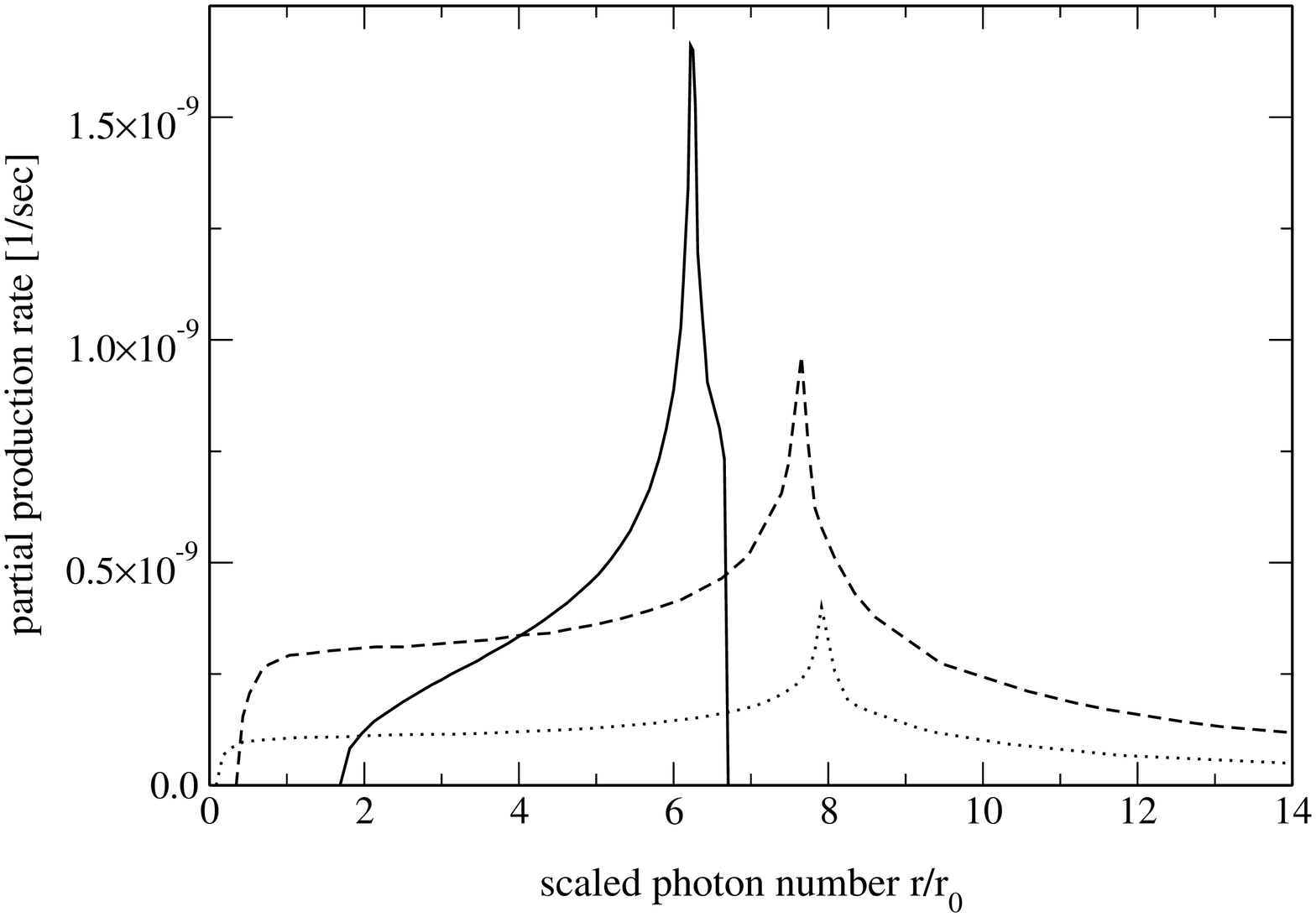}}
\end{center}
\end{figure}


\begin{thebibliography}{99}
\bibitem{hadrons} After their creation from $e^+e^-$ annihilation, the strong force 
acting between a quark and an antiquark leads to the production of more quark 
pairs, that eventually materialize into a stream of mesons and/or baryons. Therefore, 
the easiest quantity to measure experimentally is the total cross section for 
production of all hadrons.
\bibitem{notation} We employ the metric tensor $g^{\mu\nu}={\rm diag}(+---)$, 
such that the scalar product of two four-vectors $a=(a^0,\mbox{\boldmath$a$})$ 
and $b=(b^0,\mbox{\boldmath$b$})$ reads 
$(ab) = a^0b^0-\mbox{\boldmath$ab$}$, which sometimes is also denoted by 
$a\cdot b$ in order to avoid too many parentheses. Furthermore, Feynman 
slash notation is used, i.e., $\slash\!\!\!a = (\gamma a)$.
\bibitem{prop} In principle, the photon propagator is modified in the presence of 
a laser field \cite{ee1,ee2,prop2}. In the present situation, however, the free 
propagator can safely be used since the laser's electric field strength is far 
below the critical value of $E_c = m^2/e = 1.3\times 10^{16}$ V/cm.
\bibitem{res} According to Ref.\,\cite{ee+}, resonances in 
laser-assisted $e^+e^-$ scattering at low laser intensity (i.e.,
at $\xi\ll 1$) can occur only for ultrarelativistic free lepton
momenta $p_\pm$.
\bibitem{delta} The approximate treatment of the $\delta$ function can be
justified as follows. We can write the effective momentum as
$q^\mu = q^\mu(0) + \Delta q^\mu(p)$ with a $p$-independent part $q^\mu(0)$
and 
$$ \Delta q^\mu(p) \approx -\xi^2{\mbox{\boldmath$p$}^2\over 4m}{k^\mu\over\omega}. $$
Since $\Delta q^0(p)\sim \alpha_{\rm f}^2\xi^2m\ll q^0(0)$, we can perform
a Taylor expansion of the $\delta$ function, the next-to-leading-order term of 
which being proportional to $\Delta q^0(p)$. Taking this additional term into
account in the square of the reaction amplitude, results in a correction
term which is suppressed by a factor of 
$\Delta q^0(\bar p)/{\bar Q}_0\sim \alpha_{\rm f}^2\ll 1$, where $\bar p$,
${\bar Q}_0$ denote some typical values of $p$ and $Q_0$ [cf. Eq.\,(\ref{barQ})].
\bibitem{rpm} The equation $r=\beta(r)$ has two solutions: 
$r_\pm=2\xi^2(m/\omega)(1\pm \sqrt{1-\kappa^2})$. At large $r$, the 
function $J_r(\beta)$ is of significant value only in the region 
$\delta \equiv |r-\beta(r)|\sim r^{1/3}$ around $r_\pm$. The region 
is much larger around $r_+$ than around $r_-$. In fact, 
$\delta_+ \equiv |r-r_+|\approx 2\delta$ while 
$\delta_- \equiv |r-r_-|\approx \delta \cdot(M^2/2m^2\xi^2)$ for $\xi\gg M/m$.
Therefore, the main contribution to the process probability comes from 
$r\approx r_+$.
\bibitem{spread} The transversal extension of the $e^+e^-$ motion is even larger 
than a typical laser focal cross section. Further, it is worth mentioning that 
the average distance between the $e^+$ and $e^-$ moving along their classical
trajectories is larger than the spatial separation that results form the 
quantum mechanical spreading of the Ps wave packet in the laser field.
According to the classical equations of motion, the scale of the latter can be 
estimated as $\alpha_{\rm f}\xi\lambda$ (see, e.g., Ref. \cite{coll}).
\end{thebibliography}
\end{document}